\documentclass[a4paper,11pt]{article}
\pdfoutput=1 

\usepackage{jcappub} 
\usepackage{xfrac}
\usepackage{graphicx}
\usepackage{rotating}
\usepackage{physics}
\usepackage{multirow}
\usepackage[normalem]{ulem}
\usepackage{amssymb,amsmath}
\usepackage[dvipsnames]{xcolor}
\usepackage[T1]{fontenc} 

\newcommand{\lnAs}{\ln\left(10^{10} A_s\right)}
\newcommand{\Om}{\Omega_\mathrm{m}}
\newcommand{\Ob}{\Omega_\mathrm{b}}
\newcommand{\om}{\omega_\mathrm{m}}
\newcommand{\ob}{\omega_\mathrm{b}}
\newcommand{\oc}{\omega_\mathrm{cdm}}
\newcommand{\rd}{r_\mathrm{d}}
\newcommand{\Gpcoverh}{\left[ h^{-1}\mathrm{Gpc}\right]}
\newcommand{\Mpcoverh}{\left[ h^{-1}\mathrm{Mpc}\right]}

\newcommand{\hoverMpc}{\left[ h\mathrm{Mpc}^{-1} \right]}
\newcommand{\bnl}{b_{3\mathrm{nl}}}
\newcommand{\bs}{b_{s^2}}

\title{PT challenge: Validation of ShapeFit on large-volume, high-resolution mocks}

\author[1,2]{Samuel Brieden}
\author[1]{H\'ector Gil-Marín}
\author[1,3]{Licia Verde}

\affiliation[1]{ICC, University of Barcelona, IEEC-UB, Mart\'i i Franqu\`es, 1, E-08028 Barcelona, Spain}
\affiliation[2]{Dept. de F\'isica Qu\`antica i Astrof\'isica, Universitat de Barcelona, Mart\'i i Franqu\`es 1, E-
08028 Barcelona, Spain}
\affiliation[3]{ICREA, Pg. Llu\'is Companys 23, Barcelona, E-08010, Spain}

\emailAdd{sbrieden@icc.ub.edu}
\emailAdd{hectorgil@icc.ub.edu}
\emailAdd{liciaverde@icc.ub.edu }

\abstract{The ShapeFit compression method has been shown to be a powerful tool 
to gain cosmological information from galaxy power spectra in an effective, model-independent way. Here we present its performance 
on the blind PT challenge mock products presented in \cite{Nishimichi:2020tvu}.
Choosing a set-up similar to that of  other participants to the blind challenge we obtained $\Delta \lnAs  = -0.018 \pm 0.014$, $\Delta \Om = 0.0039 \pm 0.0021$ and $\Delta h =-0.0009 \pm 0.0034$, remaining below $2\sigma$ deviations for a volume of $566 \Gpcoverh^3$.
This corresponds to a  volume 10 times larger than the volume probed by future galaxy surveys. 
We also present an analysis of these mocks oriented towards an actual data analysis using the full redshift evolution, using all three redshift bins $z_1 = 0.38$, $z_2=0.51$, and $z_3 = 0.61$, and exploring different set-ups to quantify the impact of choices or assumptions on noise, bias, scale range, etc.  We find consistency across reasonable changes in set-up and across  redshifts and that, as expected, mapping the redshift evolution of clustering helps constraining cosmological parameters within a given model. }

\begin{document}
\maketitle
\flushbottom

\section{Introduction}
\label{sec:intro}
With the advent of  precision cosmology, 
our lack of understanding of the nature of the dark components of the Universe (dark matter and dark energy) has been and continues to be  the  main science driver for  galaxy redshift surveys of increasingly large cosmological volumes.  The  global effort of mapping the large scale structure (LSS)  with high fidelity and  over large cosmic volumes of the Universe continues to provide  data sets of unprecedented statistical power,  and ever more stringent constraints on cosmology \cite{BOSS:2016wmc, eBOSS:2020yzd}. It is therefore of fundamental importance to have accurate theoretical  modeling  for the key summary statistics which can then be confronted to those measured from these surveys.  As the statistical errors shrink with the larger cosmological volumes mapped, systematic errors introduced by the theoretical modeling should be kept under exquisite control. In this spirit,  the authors of \cite{Nishimichi:2020tvu} launched the (blind) PT challenge, where PT stands for ``perturbation theory''. This is a N-body mock challenge,  initially used to (blindly) test Effective Field Theory of Large Scale Structure (EFTofLSS);  the --still blind-- simulations outputs were subsequently  made available to the community and several other groups have participated to  the challenge with their own theoretical modeling approaches and implementations.

To date, all the entries to the challenge have adopted modeling methodologies that fall under the ``full modeling'' (FM) approach, in the terminology of \cite{Brieden:2021cfgLetter}. This approach follows the same philosophy as the analyses of Cosmic Microwave Background (CMB)  data: the maps are compressed into power spectrum summary statistics (a compression that is nearly lossless for CMB, and would be nearly lossless for LSS in the fully linear regime)  and these quantities are directly compared to  theory predictions for a given cosmological model to constrain the model's parameters.

But this is not the main way clustering analysis of LSS galaxy redshift surveys has been carried out for the past $\sim$15 years, which   Ref.~\cite{Brieden:2021eduPaper,Brieden:2021cfgLetter} refers to as ``classic''. In contrast to FM,  the ``classic''  clustering analysis is done by first compressing the power spectrum data into physical variables --the Alcock-Paczynski (AP) scaling factors and the amplitude of velocity fluctuations-- and then interpreting the physical variables in light of a model. The advantage of this ``classic'' approach is that the (cosmological) model-dependence is introduced at the very end of the  process, at the interpretation stage, leaving most of the analysis as model-independent as possible. It also offers a simple way to disentangle late-time physics information  from that coming from the early Universe.   The compression  aims at isolating the part of cosmological signal which information-content is least affected by systematics (see e.g., \cite{2007ApJ...665...14S, 2007ApJ...664..660E, 2004PhRvD..70j3523E, 2003ApJ...598..720S}), for this reason called BAO+RSD, where BAO stands for baryon acoustic oscillations and RSD by redshift space distortions.   This, however comes at a cost: the compression is not lossless. By emphasising robustness  and accuracy over precision, the ``classic'' compressed variables approach produces less stringent constraints on cosmological parameters than the FM approach, especially in the (minimal, flat) $\Lambda$CDM model.

Refs.~\cite{Brieden:2021eduPaper} and \cite{Brieden:2021cfgLetter} shed light on the origin, localization and physics of  extra signal responsible for the spectacular improvement in cosmological parameters constraints provided by the FM over the ``classic'' approach.   A simple, one-parameter phenomenological extension of the ``classic'' approach was hence proposed: ShapeFit. It has been shown that ShapeFit can capture most of the FM extra signal and  that it provides virtually the same statistical power in terms of cosmological parameter constraints.
In \cite{Brieden:2021cfgLetter}, ShapeFit was applied to the BOSS DR12 data and its performance  compared to that of the FM approach of a specific implementation of the EFTofLSS by \cite{Ivanov:2019pdj}.
In this paper we present the application of  ShapeFit  to the PT challenge.  While the application to real data presented in \cite{Brieden:2021cfgLetter} offer a well-rounded test,   because  real survey data include a whole suite of real-world effects, the sheer combined  volume of the PT challenge reduces the statistical errors to a point that they become negligible (its volume is an order of magnitude larger than that of the forthcoming galaxy redshift surveys \citep{desi_white_paper}) and clearly surfaces any residual modeling systematics. This is the added value of the results presented in this paper. 

The rest of the paper is organized as follows.  In section \ref{sec:data} we give a brief overview of the PT challenge set-up. In section \ref{sec:shapefit}  we review the ``classic'' and FM approaches, motivating ShapeFit and its advantages over  ``classic'' and FM; we  briefly describe the ShapeFit compression and its cosmological interpretation.   In any of these analyses several different choices  are possible in terms of priors, freedom given to nuisance parameters, and various (simulated) data cuts. Section \ref{sec:methodology} presents the different choices (set-ups) we explored and  motivates our two baseline  set-up which results  were submitted to the PT challenge. This section also includes an explanation on  how to transform (cosmological) model-independent constraints on the physical variables to constraints on the value of the cosmological parameters of a specific model (e.g., $\Lambda$CDM). The main results are presented in section \ref{sec:main_results} where we also compare the performance of ShapeFit with that of FM  in one implementation of EFTofLSS, while section \ref{sec:additional-tests} reports results for the other set-ups and serves to illustrate the sensitivity of the recovered constraints to various assumptions and modeling choices. Because the single redshift output set-up is highly idealized, in section \ref{sec:real-allz} we present results for a more realistic scenario where different redshift bins are co-analyzed.  Finally, we conclude in section \ref{sec:conclusion}.

\section{(Blind) PT challenge set-up} \label{sec:data}

The blind PT challenge was designed by \cite{Nishimichi:2020tvu} with the aim  to provide a controlled means of  testing and benchmarking theoretical models for summary statistics of galaxy redshift surveys with a particular focus on each model's or approach's  ability to recover and constrain cosmological parameter within  $\Lambda$CDM. For that purpose, they carried out a suit of simulations consisting in 10 realizations (which we will later refer to as \# 1,.., 10, each of these representing a different realization of the initial conditions at a given cosmology) in cubes of comoving side length $3840 \Mpcoverh$ with $3072^3$ particles, where the 3 input $\Lambda$CDM parameters, $\Om, A_s$ and $H_0$,  were randomly selected from a Gaussian probability distribution centered at the Planck fiducial cosmology with a width of $4\sigma$ of the Planck experiment. These randomly drawn values (which are the same for the 10 simulations) were kept secret (blind) and not known to us or any of the participants. Other cosmological parameters such as the primordial tilt and the baryon-to-matter ratio were fixed to $\Ob/\Om = 0.1571$ and $n_s = 0.9649$. 

From the simulation output, galaxy mock catalogs were generated using the \textsc{Rockstar} halo finder \citep{rockstar} and a Halo Occupation Distribution (HOD) description roughly matching BOSS galaxy data. The mock catalogs of the 10 simulations were produced for three snapshots at redshifts $z_1 = 0.38$, $z_2 = 0.51$, $z_3=0.61$ each, all coming from the evolution of the same initial conditions and therefore, not independent.

Power spectra were measured from the actual galaxy positions in redshift space. The AP distortion is later included by re-defining the $k$-vectors and the line-of-sight (LOS, constant throughout the box) assuming a fiducial cosmology with $\Om^\mathrm{fid} = 0.3$ (see section III-C of \citep{Nishimichi:2020tvu} for details). The Poissonian shot noise term was measured and subtracted from the signal. 
The authors of Ref.~\cite{Nishimichi:2020tvu} made public the measured power spectra monopole, quadrupole and hexadecapole for each simulation and each snapshot and the corresponding covariance matrix \cite{PTchallenge:data}.
The covariance matrices are provided for two different scenarios: one to match the volume and number density of BOSS data, and the other one corresponding to the simulation volume and density itself. In both cases the covariance was estimated analytically;  the correlation between different multipoles $\ell$ at same wave-vector $k$ is nonzero, and the correlation between adjacent $k$-bins is ignored. Moreover, the covariance matrices are provided individually at each snapshot and do not include any correlation among redshift bins.

Several groups have already participated in this challenge and presented their --blindly obtained-- results. Already in the initial paper by \cite{Nishimichi:2020tvu}, results were presented by the so called ``east coast'' and ``west coast'' teams who validated their implementations of the Effective Theory of Large Scale Structure (EFTofLSS) based on their works \cite{Ivanov:2019pdj} and \cite{DAmico:2019fhj}, respectively. This was followed by \cite{Chen:2020zjtPTchall} who used the PT challenge public material for testing their implementations of MONE, REPT and LPT \cite{Chen:2020fxsModel}. Finally, \cite{Kobayashi:2020zsw} participated in the challenge using a halo-model based emulator. Participants have kept the challenge blind  for the community by publishing their results (after unblinding) only as differences  between input and derived  parameters values.

\section{ShapeFit} \label{sec:shapefit}

\subsection{Motivation}\label{sec:shapefit_general_idea}

The ShapeFit methodology provides a bridge between the classic BAO+RSD approach and the FM approach: by extending the classic fixed template fit with one extra effective parameter which captures most of the information that  the classic BAO+RSD approach neglects; ShapeFit has been shown to match well  the FM constraining power while retaining the advantages of a model-agnostic compression \cite{Brieden:2021cfgLetter}.
The advantages that the ShapeFit compression offers include being:
\begin{itemize}
    \item easy to implement, because it relies on the standard BAO+RSD analysis requiring only minimal modifications to existing pipelines.
    \item easy to interpret, as the compressed parameters have an intuitive physical meaning, easy to access via analytical formulae and/or Boltzmann codes.
    \item very fast, because -as a fixed template method- it does not need calls to Boltzmann and PT codes at each evaluation step. 
    \item computationally cheap, as it only requires a minimal set of runs for a given analysis, that do not need to be repeated for each cosmological model under investigation.
    \item conveniently practical due to the reduced number of degrees of freedom (four cosmology parameters per redshift bin) with respect to the full $n$-point statistics dataset (${\cal O}(100)$). This eases massively the requirements on the number of mock catalogs needed to estimate correctly the inverse covariance matrix \cite{Hartlap:2006kj,Sellentin-Heavens,GilMarin22} in a combined pre- and post-reconstruction analysis of a real survey with overlapping redshift bins.  
    \item robust, as it makes  explicit  how each physical observable correlates with systematic uncertainties.
    \item highly modular, as the cosmological implications of each observable can be studied independently.
\end{itemize}

In what follows, we summarize the most important ingredients of ShapeFit as applied to the blind PT challenge in two steps. First, we explain the compression step based on the fixed template (or standard ruler) method and second we explain how to interpret the compressed parameters in terms of cosmological models. For a more comprehensive description we refer to Ref.~\cite{Brieden:2021eduPaper}, and in particular its figure 5 for a complete overview. 

\subsection{The ShapeFit compression} \label{sec:shapefit_compression}

The first step in the ShapeFit pipeline consists of computing the non-linear galaxy power spectrum in redshift space for a fiducial cosmology. This process is standard to most previously employed BAO+RSD or ``Full Shape'' template fits. 
The linear power spectrum template is generated with the publicly available  with CAMB \cite{Lewis:1999bs} code using the \href{https://www2.yukawa.kyoto-u.ac.jp/~takahiro.nishimichi/data/PTchallenge/data/camb_params_challenge.ini}{CAMB parameter file} provided at the PT challenge website \cite{PTchallenge:data}. Since this file does not contain information on the blinded parameters values, we choose $h^\mathrm{fid}=0.676$, $A_s^\mathrm{fid} = 2.05 \times 10^{-9}$ and $\Om^\mathrm{fid}=0.3$. The choice of $\Om^\mathrm{fid}$ is motivated by the fact that the same value is used by Ref.\cite{Nishimichi:2020tvu} for the AP distortion, while our choices for $h^\mathrm{fid}$ and $A_s^\mathrm{fid}$ are arbitrary. 

Next, we compute the perturbation theory (PT) kernels corresponding to the adopted linear template. In this work we test two PT models: one-loop Standard Perturbation Theory (1LSPT, see \cite{Beutler:2013yhm}) and two-loop Renormalized Perturbation Theory (2LRPT, see \cite{gil-marin_power_2015}). Both models incorporate four bias parameters describing the connection between dark matter and galaxy density fluctuations in real space:
the first and second order biases, $b_1$ and  $b_2$ \cite{Fry_Gazta:93}, and
the non-local biases, $b_{s2}$ and $b_\mathrm{3nl}$ \cite{McDonald_2009}. The latter are often assumed to follow the local Lagrangian relations \citep{Baldauf_2012,Saito:2014qha,kwanetal:2012}, $b_{s2}=-4/7(b_1-1)$ and $b_\mathrm{3nl}=32/315(b_1-1)$, establishing a direct link between the non-local biases and $(b_1-1)$. But these relations are motivated by theoretical considerations on the dark matter to halo connection, which does not necessarily translate equivalently into the dark matter to galaxy connection. Therefore, we may later relax this assumption in certain occasions. Our adopted redshift space formulation is based on \cite{Scoccimarro:2004tg} and extended by the TNS model \cite{Taruya:2010mx} including a Lorentzian Fingers-of-God (FoG) suppression term parametrized by $\sigma_\mathrm{FoG}$. In addition, our model allows for a shot noise term whose amplitude coefficient is  $A_\mathrm{noise}$ (see \cite{Gil-Marin:2020bct} for details) providing a correction which captures possible deviations from Poissonian statistics. 
Since shot noise is well under control in this idealized application, we employ a tight Gaussian prior $A_\mathrm{noise} = 1.00 \pm  0.01$ by default, similar to \cite{Nishimichi:2020tvu}. In a practical application, however, it is common (and recommendable) to allow for deviations from Poissonian shot noise, which can lead to additional parameter degeneracies, which, if non-Gaussian in a high-dimensional parameter space, can appear as mild biases when marginalization effectively projects them  in a lower-dimensional parameter space. Hence we also study the impact
of relaxing this prior allowing for a width of up to $30\%$, $1.00\pm0.30$. To summarize, so far we introduced six parameters per redshift bin $\left\lbrace b_1, b_2, b_{s2}, b_\mathrm{3nl}, \sigma_\mathrm{FoG}, A_\mathrm{noise} \right\rbrace$ representing our nuisance parameters $\theta_\mathrm{nuis}$ (where different priors/relations between them are possible as explained  above and as explicitly  adopted  later on).  

As common to other standard BAO+RSD template fit pipelines, the precomputed non-linear galaxy power spectrum in redshift space (plus modifications due to $\theta_\mathrm{nuis}$) is in parallel modified by a set of physical parameters per redshift bin $\theta_\mathrm{phys} = \left\lbrace \alpha_\parallel, \alpha_\perp, f, m \right\rbrace$ defined in the following. The scaling parameters $\alpha_\parallel$ and $\alpha_\perp$ add the degree of freedom of a distance dilation along and across the line of sight respectively. The growth rate $f$ allows for a variation in anisotropy that enters our adopted RSD prescription. Finally, the slope $m$ -the new ShapeFit ingredient-
is applied to the linear power spectrum \textit{a posteriori} to effectively parametrize a slope variation at the pivot scale $k_p$, at which the slope between its large scale and small scale limits reaches its maximum. In practice, this is achieved by multiplying the fiducial linear power spectrum template with the exponential of a generic sigmoid function

\begin{align} \label{eq:shapefit-m}
   {P_{\rm lin}^{\rm fid}}^\prime(k,m) = \exp \left( \frac{m}{a} \tanh{\left[a \ln\left(\frac{k}{k_p}\right) \right] } \right) \cdot P_{\rm lin}^{\rm fid}(k)
\end{align}
and replace $P_{\rm lin}^{\rm fid}(k)$ by ${P_{\rm lin}^{\rm fid}}^\prime(k)$ in each term that depends on the linear power spectrum.
As in \cite{Brieden:2021eduPaper}, we set $a=0.6$ and $k_p = 0.03 \hoverMpc$ motivated by the analytic Eisenstein and Hu, 1998 (EH98) formula \cite{Eisenstein:1997ikbaryons,Eisenstein:1997jhneutrinos}. 

\subsection{The cosmological interpretation} \label{sec:shapefit_the_cosmological_interpretation}

Let us now review the physical meaning and interpretation of the physical parameters $\theta_\mathrm{phys}$ to guide their interpretation in terms of cosmology.

It is important to keep in mind that, in the context of the fixed template approach, the scaling parameters $\alpha_\parallel, \alpha_\perp$ do not probe the absolute distance scale but rather the relative distance with respect to the standard ruler. This standard ruler is given by the sound horizon at radiation drag, $\rd$, which sets the scale of the linear power spectrum template. The  length of the standard ruler is set by  early-time physics and can be constrained by early-time physics observations when interpreted within a cosmological model for the early Universe. From late-time observations, such as LSS, and without early-time physics assumptions, the length of the standard ruler is not known. Therefore, the scalings $\alpha_\parallel, \alpha_\perp$ are interpreted as the ratio between the underlying (``true'') distances $D_\parallel, D_\perp$ and the fiducial distances $D_\parallel^\mathrm{fid}, D_\perp^\mathrm{fid}$ respectively in units of the standard ruler $\rd$, 
\begin{equation}\label{eq:shapefit_interpretation_alphas}
\begin{aligned}
\alpha_\perp(z) &= \frac{D_M(z)/\rd}{D_M^\mathrm{fid}(z)/\rd^\mathrm{fid}}~, \quad
&
\quad \alpha_\parallel(z) &= \frac{D_H(z)/\rd}{D_H^\mathrm{fid}(z)/\rd^\mathrm{fid}}~,
\end{aligned}
\end{equation}
where the distance $D_\parallel$ along the LOS is the Hubble distance $D_H(z) = c/H(z)$, the distance $D_\perp$ across the LOS is the comoving angular diameter distance $D_M(z) = \int_0^{z} c/H(z^\prime) dz^\prime$ and $H(z)$ is the Hubble expansion rate.

The growth rate $f$ is related to the logarithmic derivative of the growth factor $g(a)$ with respect to the scale factor $a$. In the context of the fixed template approach however, we need to be careful, since the fit is carried out with fixed overall amplitude. Therefore, we need to take into account that the estimated growth rate $\widetilde{f}$ -as well as the bias parameters- implicitly scales with the power spectrum amplitude. Although there are several ways to describe this, here we adopt the notation,
\begin{equation} \label{eq:shapefit_interpretation_fAmp}
    \begin{aligned}
    \widetilde{f}(z) &= \frac{d\ln g(a)}{d\ln a} A_p(z)~, \quad
    &
    \quad A_p = \left( \frac{ \left( \frac{\rd^{\rm fid}}{\rd} \right)^3  P_{\rm lin}\left( \left(\frac{\rd^{\rm fid}}{\rd}\right) k_p,z\right)}{P_{\rm lin}^{\rm fid}(k_p,z)} \right)^{1/2}~, 
    \end{aligned}
\end{equation}
where  $A_p$ is the square root of the ratio between the amplitude of the underlying power spectrum, suitably rescaled by the choice of the fiducial sound horizon, to that of the power spectrum of the fiducial cosmology at the pivot scale $k_p = \pi/\rd$. 
This convention is motivated in eqs. (3.7) and (3.8) of \cite{Brieden:2021eduPaper}. 

It is straightforward to convert $\widetilde{f}$ into the conventionally-used velocity fluctuation amplitude $f\sigma_8$. But within ShapeFit, it is convenient to define the amplitude as in eq. \eqref{eq:shapefit_interpretation_fAmp}, because at the pivot scale $k_p$ -by construction- the amplitude remains constant when varying the slope $m$. 

The slope $m$ is related to the smooth (no-wiggle) linear matter transfer function $T_{\rm nw}(k)$ which in practice we recompute at each model evaluation during posterior exploration with MCMC, and the fiducial one by 
\begin{align} \label{eq:shapefit_interpretation_m}
    m = \frac{d}{dk} \left( \ln \left[ \frac{ \left( \frac{\rd^{\rm fid}}{\rd} \right)^3 T_\mathrm{nw}^2\left( \left(\frac{\rd^{\rm fid}}{\rd}\right) k\right)}{{T_\mathrm{nw}^\mathrm{fid}}^2\left(k\right)} \right] \right) \Bigg|_{k=k_p}~.
\end{align}
where the derivative is taken at the pivot scale $k_p$.\footnote{The factor $(\rd^{\rm fid}/\rd)^3$ in the numerator of eq.~\eqref{eq:shapefit_interpretation_m} was omitted in Ref.~\cite{Brieden:2021eduPaper} as it does not change the derivative. It is included here explicitly  to make more transparent the connection to eq.~\eqref{eq:shapefit_interpretation_fAmp}.}
Following Ref.~\cite{Brieden:2021eduPaper} prescription, we use the analytic Eisenstein \& Hu 1998 (referred to as EH98) formula \cite{Eisenstein:1997ikbaryons,Eisenstein:1997jhneutrinos} to calculate the no-wiggle transfer function, but other methods like a numerical separation of the transfer function into a wiggle and a no-wiggle part are also possible (see appendix D on \cite{Briedendatapaperinprep}).

Given a cosmological model described by a set of cosmological parameters, the dependence of the physical parameters on the cosmological parameters is enclosed in $r_{\rm d}$, $D_M$, $D_H$, $g$, $P_{\rm lin}$ and $T_{\rm nw}$.

\section{Methodology} \label{sec:methodology}

We briefly describe our adopted set-up for the ShapeFit compression and its cosmological interpretation, as applied to the blind PT challenge data. We list 
several different set-up variations and motivate  our baseline choice, adopted to  represent our fiducial results which were  submitted to the blind PT challenge coordinator. 
The extensive variations on the fiducial set-up serve to quantify the robustness of the results to these choices.  

\subsection{ShapeFit compression Set-up}
\label{sec:methodology-compression}

\begin{table}
\small
\begin{tabular}{|c|c|c|c|c|c|c|c|}
\hline
Case & Bias & $\sigma_{A_\mathrm{noise}}$ & Model & $k$-range & Multipoles & Geo & Blind?  \\ \hline
\hline
S008 & $b_{\mathrm{s}^2}$, $b_{3\mathrm{nl}}$ free & $1\%$ & 1LSPT & $[0.00,0.08]$ & $\ell=0,2$ & No & Yes \\
{\bf SIM-like} & ${\boldsymbol{b_{\mathbf{s}^2}}}$\textbf{,} $\boldsymbol{ b_{3\mathbf{nl}}}$ {\bf free} &  $\boldsymbol{1\%}$ & \textbf{1LSPT} & $\boldsymbol{[0.00,0.12]}$ & $\boldsymbol{\ell=0,2}$ & \textbf{Yes} & \bf Yes \\
S1 & $b_{\mathrm{s}^2}$, $b_{3\mathrm{nl}}$ free & $ 1\%$ & 1LSPT & $[0.00,0.12]$ & $\ell=0,2$ & No & Yes \\
S2 & $b_{\mathrm{s}^2}$,  $b_{3\mathrm{nl}}$ free & $ 1\%$ & 2LRPT & $[0.00,0.12]$ & $\ell=0,2$ & No  & Yes \\
\bf DATA-like MAX & ${\boldsymbol{b_{\mathbf{s}^2}}}$\textbf{,} $\boldsymbol{ b_{3\mathbf{nl}}}$ {\bf free} & $\boldsymbol{30\%}$ & \textbf{2LRPT} & $\boldsymbol{[0.02, 0.15]}$ & $\boldsymbol{\ell=0,2,4}$ & \textbf{No} &  {\bf Yes} \\
\bf " (geo)  & ${\boldsymbol{b_{\mathbf{s}^2}}}$\textbf{,} $\boldsymbol{ b_{3\mathbf{nl}}}$ {\bf free} & $\boldsymbol{30\%}$ & \textbf{2LRPT} & $\boldsymbol{[0.02, 0.15]}$ & $\boldsymbol{\ell=0,2,4}$ & \textbf{Yes} &  \bf No \\
Dbs2b3nl-20 & $b_{\mathrm{s}^2}$,  $b_{3\mathrm{nl}}$ free & $20\%$ & 2LRPT & $[0.02, 0.15]$ & $\ell=0,2,4$ &  No &  Yes \\
Dbs2b3nl-15 & $b_{\mathrm{s}^2}$,  $b_{3\mathrm{nl}}$ free & $15\%$ & 2LRPT & $[0.02, 0.15]$ & $\ell=0,2,4$ &  No  & Yes \\
Dbs2b3nl-5 & $b_{\mathrm{s}^2}$,  $b_{3\mathrm{nl}}$ free & $5\%$ & 2LRPT & $[0.02, 0.15]$ & $\ell=0,2,4$ &  No  & Yes \\
Dbs2b3nl-1 & $b_{\mathrm{s}^2}$,  $b_{3\mathrm{nl}}$ free & $1\%$ & 2LRPT & $[0.02, 0.15]$ & $\ell=0,2,4$ &  No  &  Yes \\
Dbs2b3nl-1-geo & $b_{\mathrm{s}^2}$,  $b_{3\mathrm{nl}}$ free & $1\%$ & 2LRPT & $[0.02, 0.15]$ & $\ell=0,2,4$ &  Yes  & Yes  \\
Dbs2b3nl-1-12 & $b_{\mathrm{s}^2}$,  $b_{3\mathrm{nl}}$ free & $1\%$ & 2LRPT & $[0.02, 0.12]$ & $\ell=0,2,4$ & No &  Yes \\
Dbs2b3nl-1-20 & $b_{\mathrm{s}^2}$,  $b_{3\mathrm{nl}}$ free & $1\%$ & 2LRPT & $[0.02, 0.20]$ & $\ell=0,2,4$ & No &  Yes \\
Dbs2b3nl-1-25 & $b_{\mathrm{s}^2}$,  $b_{3\mathrm{nl}}$ free & $1\%$ & 2LRPT & $[0.02, 0.25]$ & $\ell=0,2,4$ & No &  Yes \\
Dbs2b3nl-1-12-geo & $b_{\mathrm{s}^2}$,  $b_{3\mathrm{nl}}$ free & $1\%$ & 2LRPT & $[0.02, 0.12]$ & $\ell=0,2,4$ & Yes &  No \\
Dbs2b3nl-1-20-geo & $b_{\mathrm{s}^2}$,  $b_{3\mathrm{nl}}$ free & $1\%$ & 2LRPT & $[0.02, 0.20]$ & $\ell=0,2,4$ & Yes &  No \\
Dbs2b3nl-1-25-geo & $b_{\mathrm{s}^2}$,  $b_{3\mathrm{nl}}$ free & $1\%$ & 2LRPT & $[0.02, 0.25]$ & $\ell=0,2,4$ & Yes &  No \\
Dbs2-30 & $b_{\mathrm{s}^2}$ free & $30\%$ & 2LRPT & $[0.02, 0.15]$ & $\ell=0,2,4$ &  No & Yes \\
Dbs2-30-geo & $b_{\mathrm{s}^2}$ free & $30\%$ & 2LRPT & $[0.02, 0.15]$ & $\ell=0,2,4$ & Yes & No \\
Dbs2-20 & $b_{\mathrm{s}^2}$ free & $20\%$ & 2LRPT & $[0.02, 0.15]$ & $\ell=0,2,4$ &  No &  Yes \\
Dbs2-15 & $b_{\mathrm{s}^2}$ free & $15\%$ & 2LRPT & $[0.02, 0.15]$ & $\ell=0,2,4$ &  No &  Yes \\
Dbs2-5 & $b_{\mathrm{s}^2}$ free & $5\%$ & 2LRPT & $[0.02, 0.15]$ & $\ell=0,2,4$ &  No &  Yes \\
Dbs2-1 & $b_{\mathrm{s}^2}$ free & $1\%$ & 2LRPT & $[0.02, 0.15]$ & $\ell=0,2,4$ &  No &  Yes \\
Dbs2-1-geo & $b_{\mathrm{s}^2}$ free & $1\%$ & 2LRPT & $[0.02, 0.15]$ & $\ell=0,2,4$ &  Yes & Yes \\
Db3nl-30 & $b_{3\mathrm{nl}}$ free & $30\%$ & 2LRPT & $[0.02, 0.15]$ & $\ell=0,2,4$ &  No & Yes \\
Db3nl-30-geo & $b_{3\mathrm{nl}}$ free & $30\%$ & LRPT & $[0.02, 0.15]$ & $\ell=0,2,4$ &  Yes & No \\
Db3nl-20 & $b_{3\mathrm{nl}}$ free & $20\%$ & 2LRPT & $[0.02, 0.15]$ & $\ell=0,2,4$ &  No & Yes \\
Db3nl-15 & $b_{3\mathrm{nl}}$ free & $15\%$ & 2LRPT & $[0.02, 0.15]$ & $\ell=0,2,4$ &  No & Yes \\
Db3nl-5 & $b_{3\mathrm{nl}}$ free & $5\%$ & 2LRPT & $[0.02, 0.15]$ & $\ell=0,2,4$ &  No & Yes \\
Db3nl-1 & $b_{3\mathrm{nl}}$ free & $1\%$ & 2LRPT & $[0.02, 0.15]$ & $\ell=0,2,4$ &  No & Yes \\
Db3nl-1-geo & $b_{3\mathrm{nl}}$ free & $1\%$ & 2LRPT & $[0.02, 0.15]$ & $\ell=0,2,4$ &  Yes & Yes \\
{\bf DATA-like MIN} & \textbf{all local} & $\boldsymbol{30\%}$ & \textbf{2LRPT} & $\boldsymbol{[0.02, 0.15]}$ & $\boldsymbol{\ell=0,2,4}$ &  \textbf{No} & \bf Yes \\
\bf " (geo) & \textbf{all local} & $\boldsymbol{30\%}$ & \textbf{2LRPT} & $\boldsymbol{[0.02, 0.15]}$ & $\boldsymbol{\ell=0,2,4}$ &  \textbf{Yes} & \bf No \\
Dblocal-20 & all local & $20\%$ & 2LRPT & $[0.02, 0.15]$ & $\ell=0,2,4$ &  No & Yes \\
Dblocal-15 & all local & $15\%$ & 2LRPT & $[0.02, 0.15]$ & $\ell=0,2,4$ &  No & Yes \\
Dblocal-5 & all local & $5\%$ & 2LRPT & $[0.02, 0.15]$ & $\ell=0,2,4$ &  No & Yes \\
Dblocal-1 & all local & $1\%$ & 2LRPT & $[0.02, 0.15]$ & $\ell=0,2,4$ &  No & Yes \\
Dblocal-1-geo & all local & $1\%$ & 2LRPT & $[0.02, 0.15]$ & $\ell=0,2,4$ &  Yes  & Yes \\
\hline
\end{tabular}
\caption{ Blind PT challenge set-up overview. Most fits were carried out before unblinding, but some of them (indicated by the final column) were updated after unblinding to take into account the geometric correction, see text for more details. Only the boldface cases were submitted to the blind PT challenge coordinator including the DATA-like runs corrected after unblinding. These are discussed in section \ref{sec:main_results}. The other cases represent further tests of the different ingredients such as the bias option, the prior on shot noise, the PT model, the fitted scale range and the geometric correction (Geo) presented in section \ref{sec:additional-tests}. Unless specified otherwise in the second column, each of the non-local bias parameters $b_{\mathrm{s}^2}$,  $b_{3\mathrm{nl}}$ is set to the local Lagrangian prediction. In all cases we fit the mean of all 10 realizations at the highest redshift bin $z_3 = 0.61$ using the corresponding high-volume covariance matrix. Not all these cases are discussed in details in the main text or shown in the figures, but are presented here for completeness. }\label{tab:methodology_all_set-ups}
\end{table}

For our baseline analysis we take the average of the power spectra of all 10 realisations at a single redshift, in this particular case we chose $z_3 = 0.61$. We also analyzed the $z_1$ and $z_2$ redshift bins, finding equivalent results.
The provided covariance, corresponding to the volume of a single realization, was rescaled to a volume of $566 \Gpcoverh$ (10 realisations).

As explained in section \ref{sec:shapefit_compression}, we create a template using the blind PT challenge CAMB input file, adding $h^\mathrm{fid}=0.676$, $A_s^\mathrm{fid} = 2.05 \times 10^{-9}$ and $\Om^\mathrm{fid}=0.3$. We compute the first and second-order loop corrections using \textsc{PTcool} \cite{GilMarin:PTcool}.
The ShapeFit analysis is carried out by varying the four physical parameters,  $\theta_{\rm phys}$, and several combinations of nuisance parameters. We consider different fitting configurations exploring combinations of nuisance parameters being varied, multipoles and range of scales  considered, modelling of non-linearities, which are all listed in table \ref{tab:methodology_all_set-ups} \footnote{Not all the cases presented in the table are discussed in details in the main text or shown  in the figures, but are presented in the table  for completeness. In the spirit of open science, readers interested in the outputs of specific runs can send a reasonable request to the authors.}.  Broadly, these set-ups can be split in two categories: 
\begin{enumerate}
    \item {\bf SIM-like} (also labeled as initial letter ``S''). 
    These set-ups
    are tuned towards fitting a synthetic, simulated  and thus idealized dataset, where the very large scales are under control and the shot noise is known. Therefore, we employ a narrow $1\%$ prior on the shot noise amplitude $A_\mathrm{noise}$ and set $k_\mathrm{min} = 0.0$, not imposing any large scale cut. This is compatible with the analysis choices of other blind PT challenge participants. To be consistent with the choices of  most of the other participants, we also set $k_\mathrm{max} = 0.12 \hoverMpc$, fit the monopole and quadrupole only, and choose 1LSPT as our baseline modeling for non-linearities.
    \item {\bf DATA-like} (also labeled as initial letter ``D''). These cases are tuned towards an actual data analysis, similar to the methodologies employed in \cite{BOSS:2016wmc,eBOSS:2020yzd}, where large scales are affected by systematics and the shot noise value is unknown. Therefore, we allow for a broader prior on $A_\mathrm{noise}$ of up to $30\%$ and set a large-scale cut at $k_\mathrm{min} = 0.02 \hoverMpc$. As in the data analyses of \cite{BOSS:2016wmc,eBOSS:2020yzd} we set $k_\mathrm{max} = 0.15 \hoverMpc$, fit monopole, quadrupole and hexadecapole and choose the 2LRPT as baseline modeling for non-linearities for our template-fits.
\end{enumerate}

As far as non-local bias parameters $b_{\mathrm{s}^2}$ and $b_{3\mathrm{nl}}$ are concerned, for the DATA-like cases we explore all possible combinations of setting them to their local Lagrangian prediction (``local'' in the label)  or varying them freely. For the SIM-like cases, we vary both parameters freely,
to explore potential deviations from the local Lagrangian dark matter-halo connection and to enable a more direct comparison with previous analyses based on the EFT approach. For more details, see the section 4.2 of \cite{Brieden:2021eduPaper}. There
it is shown that varying $b_{\mathrm{s}^2}$ and $b_{3\mathrm{nl}}$ (labeled ``MAX'' in \cite{Brieden:2021eduPaper}) is equivalent to the EFT case with varying $b_{\mathrm{s}^2}$ and counterterm $c_0$.
In fact, $c_0$ has been found to be very degenerate with $b_{3\mathrm{nl}}$ \cite{Ivanov:2019pdj,Nishimichi:2020tvu}, hence varying only one of the two parameters is equivalent to varying only the other (or varying both).
In \cite{BOSS:2016wmc,eBOSS:2020yzd} on the other hand,  both bias parameters are set to their local Lagrangian predictions (labeled ``MIN'' in \cite{Brieden:2021eduPaper}). This choice has been shown to be valid for the BOSS and eBOSS analyses which focus on BAO scales, neglecting large-scales broadband shape information \cite{Gil-Marin:2020bct, Briedendatapaperinprep}. As it is not clear yet whether this strategy remains valid for upcoming surveys, we submitted to the blind challenge both the freely varying (DATA-like MAX) and local Lagrangian (DATA-like MIN) cases. For the submitted SIM-like case, it is important to note,  we employed the ``geometric correction'' introduced in section 6 of \cite{Brieden:2021eduPaper}.
At the largest scales, where $k$-bins are broad and there are few modes per bin,  mode-discreteness introduces a mismatch between the conventionally averaged and the mode-averaged model evaluated at each bin. The ``geometric correction'' is a fast way to account for this effect.
As we show in figure \ref{fig:main-results-tests1}, this correction is important when fitting the whole $k$-range,
and even when applying a large-scale cut at $k_\mathrm{min} = 0.02 \hoverMpc$.\footnote{We realized this fact after unblinding as explained in the second paragraph of section \ref{sec:main_results}.}

The adopted parameters priors for the fits are given in  the second and third sections of table \ref{tab:methodology-priors}. 
The results of the three submitted fits (highlighted in boldface in table \ref{tab:methodology_all_set-ups}) are presented in section \ref{sec:main_results}. The remaining, not highlighted, cases are presented in section \ref{sec:additional-tests}, where further tests of the different configuration ingredients such as the bias option, the prior on shot noise, the PT choices in modeling of non-linearities, the fitted scale range and the geometric correction are discussed.

\begin{table}[t]
    \centering
    \begin{tabular}{|c|c|c|c|}
      \hline 
       \multicolumn{2}{|c}{Parameter} & \multicolumn{2}{|c|}{Prior} \\ \hline
       type & name & range & type \\ \hline
       \hline
     & $\omega_\mathrm{b}$  &  $[\mathrm{0.005}, \mathrm{0.04}]$ & flat \\
     & $\omega_\mathrm{cdm}$  &  $[\mathrm{0.01}, \mathrm{0.99}]$ & flat \\
      \multirow{2}{*}{Cosmological} & $h$  &  $[0.1, 3.0]$ & flat \\
     \multirow{2}{*}{($\Lambda$CDM Fit)} & $\ln\left(10^{10} A_s\right)$  & $[0.1, 10]$ & flat \\
      & $\omega_\mathrm{b}/(\omega_\mathrm{b}+\omega_\mathrm{cdm})$  &  $(0.1571 \pm 0.0001)$ & Gaussian \\
      & $n_s$  &  $0.9649$ & fixed \\
      & $M_\nu\,[\mathrm{eV}]$  &  $0.0$ & fixed \\
      \hline
      & $\alpha_\parallel$  & $[0.5, 1.5]$ & flat \\
    Physical & $\alpha_\perp$ & $[0.5, 1.5]$ & flat \\
  (ShapeFit)  & $f$ & $[0, 3]$ & flat  \\
     & $m$ & $[-3, 3]$ & flat  \\ \hline
     & $b_1$ & $[0, 20]$ & flat \\
    & $b_2$ & $[\text{-} 20, 20]$ & flat \\
    Nuisance    & $b_{s2}$ & lag. / $[\text{-} 20, 20]$ & derived / flat \\
    (ShapeFit) & $b_{3\mathrm{nl}}$ & lag.  / $[\text{-} 20, 20]$ & derived / flat \\
     & $\sigma_P \, \Mpcoverh$ & $[0, 10]$ & flat \\
     & $A_\mathrm{noise}$ & $(1 \pm \sigma_{A_\mathrm{noise}})$ & Gaussian \\
     \hline
    \end{tabular}
    \caption{Prior ranges for parameters used for \textit{ShapeFit} and the cosmological fit. Flat priors are given as $[\mathrm{min},\mathrm{max}]$, Gaussian priors are denoted as $(\mathrm{mean} \pm \mathrm{std})$. The prior knowledge about the baryon fraction $\ob/\om = 0.1571$ is implemented as a Gaussian prior with very small width, which is equivalent to treating the ratio as fixed.}
    \label{tab:methodology-priors}
\end{table}

\subsection{From physical parameters  to cosmological parameters constraints (and back)} \label{sec:methodology-cosmlogy}

While for any set of values for the cosmological parameters of a given cosmological model it is always possible to derive the corresponding physical parameter values, the converse is not true. There can be regions of physical parameter space that do not correspond to any choice of cosmological parameters within the families of models under consideration.
To convert the ShapeFit constraints on the physical parameters into  constraints on cosmological parameters for a given cosmological model we proceed as described in section \ref{sec:shapefit_the_cosmological_interpretation}. 

For the fiducial model we compute the quantities $\rd^\mathrm{fid}, D_M^\mathrm{fid}(z), D_H^\mathrm{fid}(z), P_\mathrm{lin}^\mathrm{fid} (k_p,z)$ and the fiducial smooth EH98 power spectrum.
In-line with the blind PT challenge recommendations, we vary $\left\lbrace \ob, \oc, h, \lnAs \right\rbrace$ while setting the neutrino mass to zero, fixing the scalar tilt $n_s = 0.9649$ and employing a tight Gaussian prior on the baryon to matter ratio $\ob/\om = 0.1571\pm 0.0001$. This prior basically fixes $\ob$ for a given input value of $\oc$, such that the degrees of freedom of our $\Lambda$CDM fit are $\left\lbrace \oc, h, \lnAs \right\rbrace$.  With these choices we are matching the set-up of the other blind PT challenge participants.
For each choice of cosmological parameters, and with the pre-computed values of the fiducial quantities, we obtain model predictions for $\left\lbrace \alpha_\parallel(z), \alpha_\perp(z), f(z) A_p(z), m \right\rbrace$ which are then compared with the ShapeFit constraints obtained from a cosmological MCMC using MontePython
\cite{Brinckmann:2018cvx,Audren:2012wb,Brinckmann:MontePythonCode}. But note that running additional chains is actually not necessary, as they can instead be easily obtained from the ShapeFit chains via importance sampling \cite{importance_sampling}.  
This way of proceeding effectively imposes a $\Lambda$CDM prior on the ShapeFit results. With this approach it is then possible to visualize  what the ShapeFit  physical parameters constraints would be in a $\Lambda$CDM model by recomputing  the physical parameters as if they were derived parameters of the ($\Lambda$CDM) cosmological parameters.  

A summary of all parameters and priors used for ShapeFit and for the cosmological $\Lambda$CDM fit is given by table \ref{tab:methodology-priors}.

\section{Results submitted to the blind PT challenge and changes post-unblinding} \label{sec:main_results}

The results initially presented in this section (but now slightly updated as described below) were submitted to the blind PT challenge coordinator.
After submission, the organizer revealed to us the true underlying model values (unblinding). To keep the PT challenge blind for the rest of the community  in all figures and tables throughout this work we only show our results as differences from the true values.

More specifically, for any measured parameter $\theta$ we show $\Delta \theta = \theta - \theta^\mathrm{true}$. This applies both to physical and cosmological parameter constraints.
Results not submitted to the PT challenge coordinator (some computed before the unblinding, see table \ref{tab:methodology_all_set-ups}) are presented in section~\ref{sec:additional-tests}.

We initially estimated that the geometric correction would be unimportant if a minimum wavenumber $k$ cut was imposed. This is certainly true for current and forthcoming surveys, for the full PT challenge volume it is still not statistically significant --not applying the geometric correction induces  small shifts in the recovered variables, less than  1-$\sigma$-- but the effect is visible under close inspection.  The results initially submitted neglected the geometric correction, which was then included subsequently. All the results presented here always include the geometric correction even with a minimum $k$ cut. As a consequence, these results are technically not blind (see table \ref{tab:methodology_all_set-ups} for more details) but the changes compared to the blind version can only be appreciated upon very careful scrutiny.
\begin{figure}[t]
    \centering
    \includegraphics[scale=0.45]{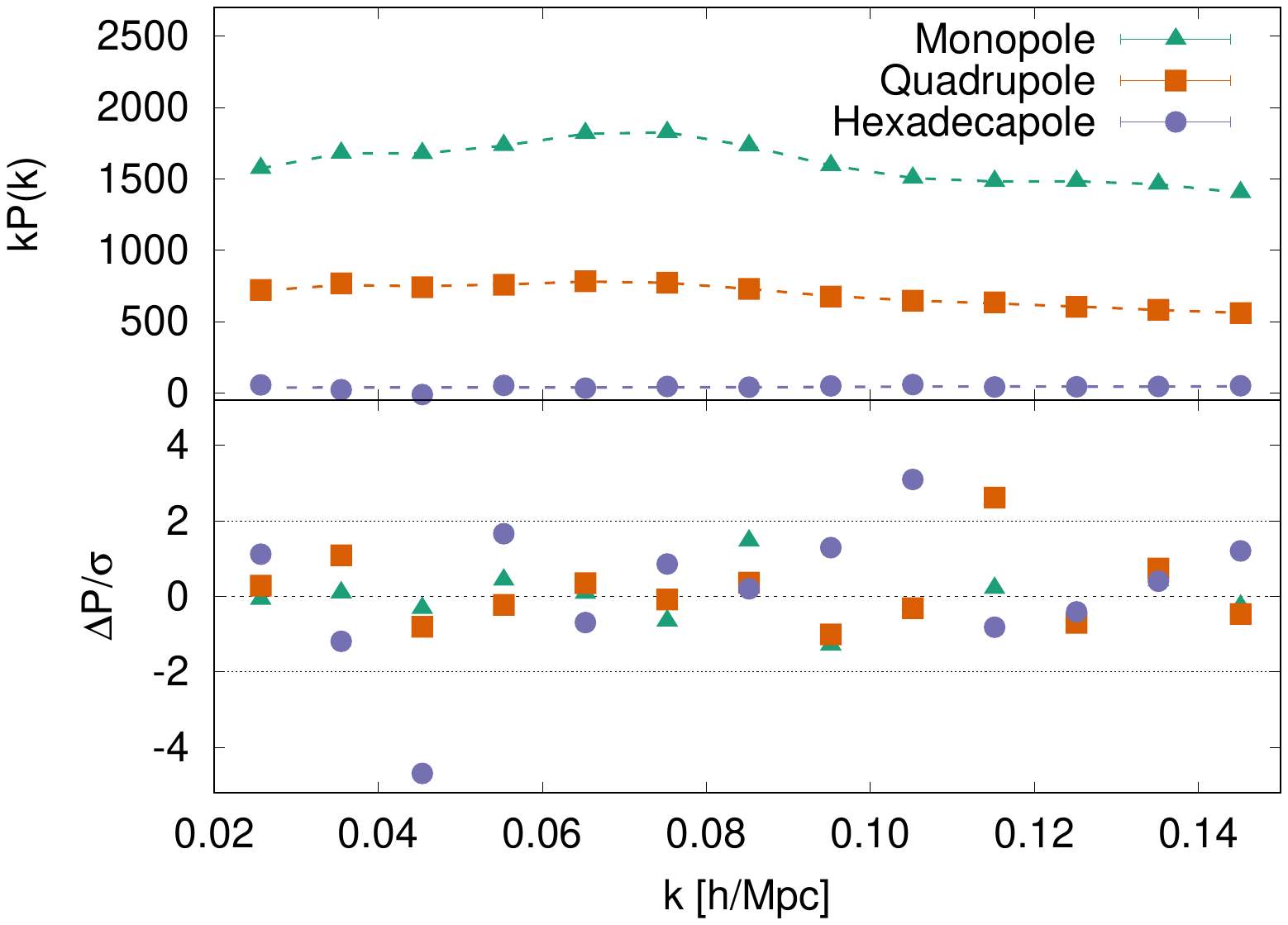}
    \caption{This figure shows the simulation data versus model comparison for the DATA-like MAX case. In the upper panel we show the multipole power spectra data and bestfit model and in the bottom panel their residuals in unit of the errors corresponding to a   volume of $566 \Gpcoverh^3$. The monopole, quadrupole and hexadecapole data and bestfit are displayed as described in the legend.
    }
    \label{fig:model_data_comp}
\end{figure}

The ShapeFit bestfit to the PT challenge multipoles is shown for one particular configuration (DATA-like MAX) in the top panel of figure \ref{fig:model_data_comp}. Since the error bars correspond to the mean of the realizations equivalent to $566\Gpcoverh$ they are not visible by eye, but from the residuals in the lower panel it can be appreciated that the model delivers a good fit on the displayed wavevector range up to $k_\mathrm{max} = 0.15 \hoverMpc$.  The data points mostly remain within the $2\sigma$ band, but for the hexadecapole there is visibly a larger scatter, the most  divergent ($4.5 \sigma$) data point is at $k=0.045 \hoverMpc$. This hints towards a problem with the hexadecapole related to the inhomogeneous distribution of the directions of the $k$-vectors, that we will return to later.

\begin{figure}[b]
    \centering
    \begin{minipage}{0.56\textwidth}
    \includegraphics[width=\textwidth]{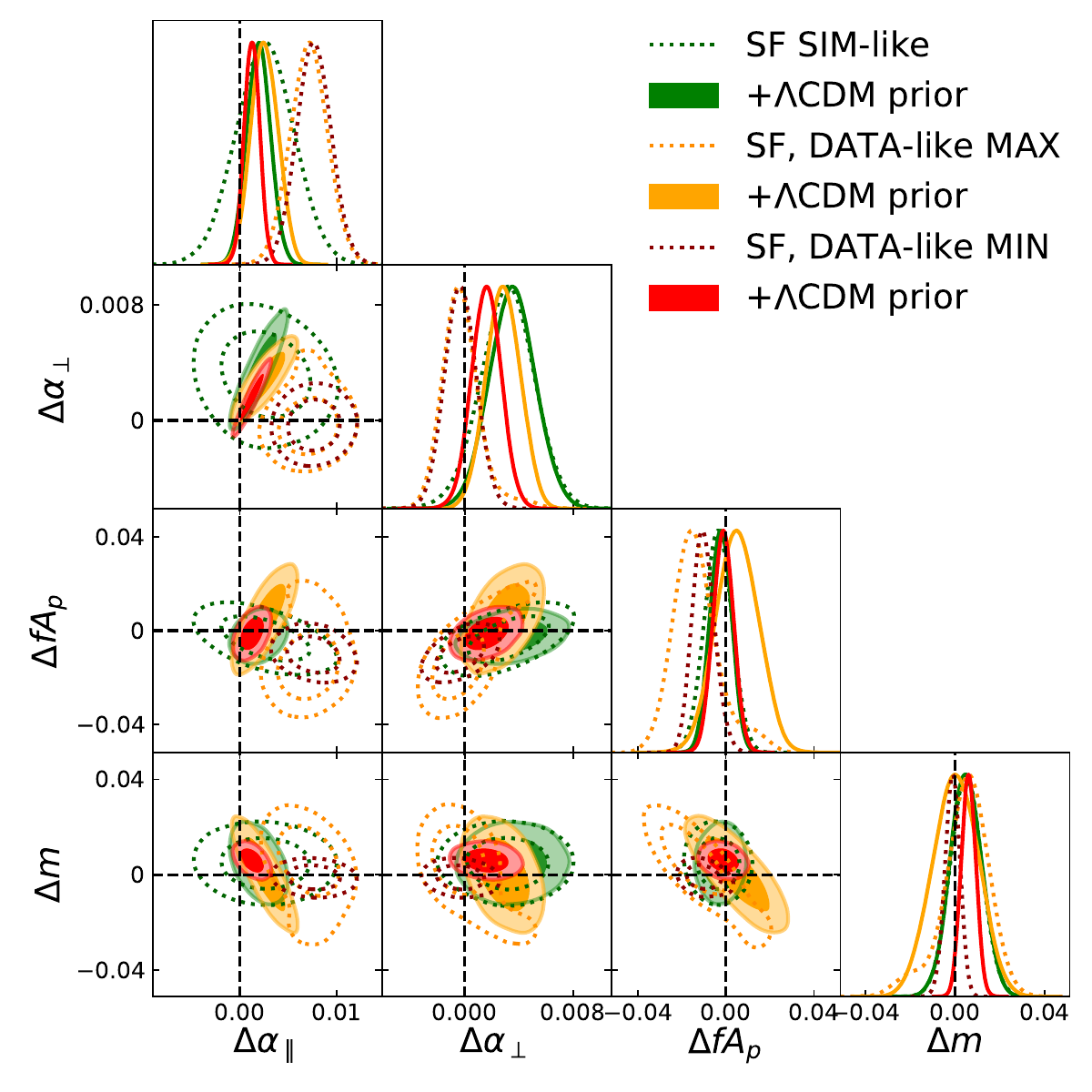}
    \end{minipage}
    \begin{minipage}{0.43\textwidth}
    \vspace{1.8cm}
    \includegraphics[width=\textwidth]{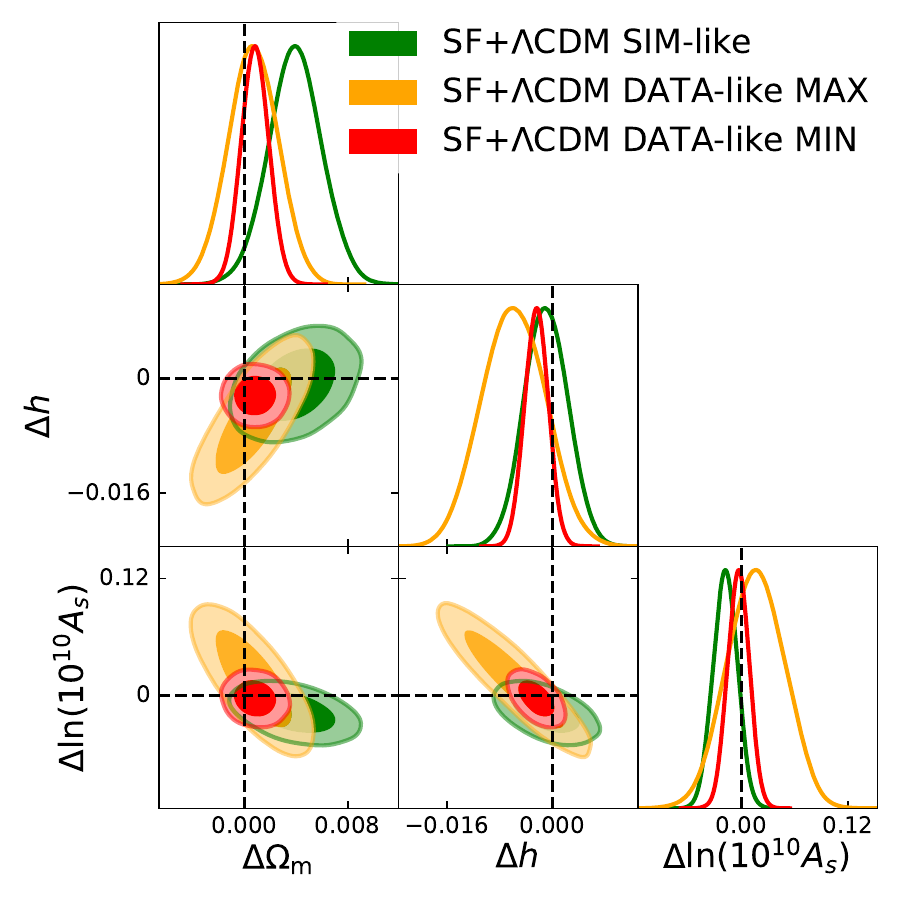}
    \end{minipage}
    \caption{These represent our main results for the blinded physical parameters (left panel) and the blinded cosmological parameters $\left\lbrace \Delta h, \Delta \Om, \Delta\lnAs \right\rbrace$ (right panel). To ensure that the PT challenge remains blind for the rest of the community, results are visualized as differences from their true values. The dashed vertical and horizontal lines guide the eye to zero difference. Dotted, empty contours represent the ShapeFit (SF) results and continuous, filled contours the corresponding (cosmological parameters) results with a $\Lambda$CDM prior imposed. We show the SIM-like results in green, the DATA-like MAX case in orange and the DATA-like MIN case in red. The numerical results can also be found in table \ref{tab:main_results}.}
    \label{fig:main-results}
\end{figure}

The physical parameters constrained with ShapeFit are shown as dashed curves in the left panel of figure \ref{fig:main-results}. The SIM-like constraints (see specifications in table \ref{tab:methodology_all_set-ups}) are displayed in green-dotted, the DATA-like MAX and MIN cases in orange and red-dotted, respectively. For all cases we recover the true values $\Delta \theta_\mathrm{phys}=0$ (indicated by black dashed lines) well within $1.5\sigma$,  except for $\alpha_\parallel$ in the DATA-like cases,  where  we find a bias of $4-5\sigma$.
In the DATA-like cases there are four main changes compared to the SIM-like cases: use of 2LRPT instead of 1LSPT, the wider prior on the shot-noise amplitude, the higher $k_{\rm max}$ and the inclusion of the hexadecapole. We verify in section \ref{sec:additional-tests}
that the inclusion of the hexadecapole drives the systematic shift; in particular, it shifts $\alpha_\parallel$ by $\Delta_\alpha = 0.005$, which amounts to most\footnote{The remaining bit of $\Delta_\alpha = 0.003$ can be explained by the choice of maximum wavevector. See section \ref{sec:additional-tests} and figure \ref{fig:main-results-tests-kmax} in particular.} of the difference between the maximum of the green-dotted and range-dotted $\alpha_\parallel$ posteriors. The effect on the $\alpha$'s of the inclusion of the hexadecapole has not been explored thoroughly in the literature (but see \citep{Gil-Marin:2020bct} and footnote in sec B1 of \cite{Nishimichi:2020tvu}). The shift induced by the hexadecapole  seen here is below the $2\sigma$ level for DESI-like survey and not significant for a BOSS/eBOSS-like survey.  We leave this to further investigation in future work.
Also, the higher $k_{\rm max}$ and the inclusion of the hexadecapole result in smaller error-bars for the $\alpha$s. The size of the error-bars on $m$ (and somewhat on $fA_p$) on the other hand is driven by the choice of prior on the nuisance bias parameters. 

The ShapeFit results, which are  (cosmological) model-independent, can be interpreted within the $\Lambda$CDM model as discussed in section~\ref{sec:methodology-cosmlogy}. We refer to this as  imposing a $\Lambda$CDM prior and this case is shown as filled contours in both panels of figure \ref{fig:main-results}. It is easier to understand the filled contours in the left panel of figure \ref{fig:main-results} after considering the right figure panel, where the ShapeFit+$\Lambda$CDM prior constraints are shown for the cosmological parameters constrained by the analysis: $h$, $\lnAs$, $\Omega_m$. 
For all cases we find excellent agreement with the true cosmology with biases of at most $2\sigma$.
Comparing the DATA-like MIN and MAX cases, we find that the local Lagrangian bias assumption leads to reduced error bars but does not bias the cosmological results, even given the precision of the PT challenge suite of simulations.
Recall that the error-bars reported here for the PT challenge are for a volume of $566\, {\rm Gpc}{h}^{-1}$ which is about 10 times bigger than the volume covered by forthcoming surveys. In section \ref{sec:additional-tests} we further investigate the various bias assumptions and discuss their relevance for the ongoing and future surveys. 

The $\lesssim 2\sigma$ deviations of the DATA-like cases in the cosmological parameter space seem mild compared to the $4\sigma$ deviations seen in the physical parameter space. To understand  this, in the left panel of figure \ref{fig:main-results} we show  the physical parameters constraints corresponding to the $\Lambda$CDM parameter constraints of the right panel.
The filled contours in the left panel are hence derived from the dashed, empty contours by imposing a $\Lambda$CDM prior.

In the SIM-like case (green contours) this prior does not affect much $\alpha_\perp$, $fA_p$ and $m$, but significantly tightens the $\alpha_\parallel$ posterior. The reason is that within $\Lambda$CDM there is a tight correlation between $\alpha_\parallel$ and $\alpha_\perp$. By construction they are related for any model, because the perpendicular distance is the integral of the parallel distance (see the definitions below eq. \eqref{eq:shapefit_interpretation_alphas}). Within $\Lambda$CDM however, the redshift evolution of the Hubble expansion rate is completely determined by $\Omega_\mathrm{m}$, meaning that for a given $\Omega_\mathrm{m}$, $\alpha_\parallel$ and $\alpha_\perp$ are not independent. 
In the ShapeFit compression with a $\Lambda$CDM prior, $\Omega_\mathrm{m}$ is constrained by $m$, hence $\alpha_\parallel$ and $\alpha_\perp$ become 
directly linked. 

\begin{table}[t]
\centering
\begin{tabular}{|c|c|c|c|c|}
\hline
Type & Parameter & SIM-like & DATA-like MAX & DATA-like MIN \\ \hline \hline
\multirow{4}{*}{Physical} & $\Delta \alpha_\parallel$ & $0.0026 \pm 0.0030$ & $0.0071 \pm 0.0019$ & $0.0076 \pm 0.0018$ \\
& $\Delta \alpha_\perp$ & $0.0031 \pm 0.0020$ & $-0.0002 \pm 0.0015$ & $-0.0003 \pm 0.0012$ \\ 
& $\Delta f A_p$ & $-0.0035 \pm 0.0063$ & $-0.0131 \pm 0.0111$ & $-0.0101 \pm 0.0052$
\\
& $\Delta m$ & $0.0049 \pm 0.0071$ & $0.0042 \pm 0.0112$ & $-0.0012 \pm 0.0034$ \\  \hline \hline
\multirow{2}{*}{$\Lambda$CDM} & $\Delta \Omega_\mathrm{m}$ & $0.0039 \pm 0.0021$ & $0.0006 \pm 0.0020$ & $0.0008 \pm 0.0011$ \\
\multirow{2}{*}{Cosmological}
& $\Delta h$ & $-0.0009 \pm 0.0034$ & $-0.0059 \pm 0.0048$ & $-0.0025 \pm 0.0018$ \\ 
& $\Delta \ln (10^{10} A_s)$ & $-0.0182 \pm 0.0136$ & $0.0168 \pm 0.0325$ & $-0.0032 \pm 0.0121$ \\
\hline \hline
 & $\Delta \alpha_\parallel$ & $0.0019 \pm 0.0012$ & $0.0025 \pm 0.0014$ & $0.0013 \pm 0.0008$ \\
$\Lambda$CDM & $\Delta \alpha_\perp$ & $0.0035 \pm 0.0017$ & $0.0028 \pm 0.0012$ & $0.0016 \pm 0.0011$ \\ 
Derived & $\Delta f A_p$ & $-0.0024 \pm 0.0048$ & $0.0053 \pm 0.0097$ & $-0.0012 \pm 0.0046$ \\
& $\Delta m$ & $0.0048 \pm 0.0071$ & $-0.0084 \pm 0.0104$ & $-0.0026 \pm 0.0034$ \\
\hline
\end{tabular}
\caption{Here we show the mean values and their corresponding symmetrized errorbars of the constrained parameters for the SIM-like, DATA-like MAX and DATA-like MIN cases specified in table \ref{tab:methodology_all_set-ups}. Using the configuration of table \ref{tab:methodology-priors}, the physical parameters are constrained with ShapeFit and the cosmological parameters are obtained by fitting the $\Lambda$CDM model to the ShapeFit results. Finally, we show again the physical parameters but this time derived from the cosmological fits, hence with a $\Lambda$CDM prior. The results shown here correspond to the same results shown in figure \ref{fig:main-results}.}
\label{tab:main_results}
\end{table}

Considering the DATA-like cases we notice that -under the umbrella of $\Lambda$CDM- their $\alpha_\parallel$ posteriors agree very well with the SIM-like case, even though the pure ShapeFit constraints show a $4\sigma$ deviation. The filled contours are markedly shifted towards the true parameter values compared to the dashed contours.  This means that the bias in $\alpha_\parallel$ observed in the ShapeFit DATA-like cases displaces $\alpha_\parallel$ to a region which is excluded by (or unphysical in)  $\Lambda$CDM. Hence, the $\Lambda$CDM prior drives $\alpha_\parallel$ towards its correct value, at the expense of creating a tension between the $\Lambda$CDM and the model-independent result for $\alpha_\parallel$. 
For the other physical parameters the shifts induced by imposing the $\Lambda$CDM are much less dramatic.

As further explored in sec.~\ref{sec:additional-tests}, it is important to investigate the ShapeFit model ingredients, such as the bias assumption, the shot noise prior, and others which may be responsible for the $\alpha_\parallel$ systematic bias. Nevertheless the left panel of figure \ref{fig:main-results} highlights  an important point:  some systematic biases can only be seen at the  model-independent parameter compression stage and would not be spotted in the context of direct, model-dependent fits (especially for minimal-$\Lambda$CDM type-models).  

Our main results are also summarized in table \ref{tab:main_results}, where we show the physical parameters constrained with ShapeFit, the cosmological results of the $\Lambda$CDM fits and again the physical parameters, but this time derived from the model (which is equivalent to imposing a $\Lambda$CDM prior). As discussed before, the agreement with the truth is very good, especially for the SIM-like case. In the DATA-like cases, there are notable biases in $\alpha_\parallel$, that vanish once they are interpreted in light of the $\Lambda$CDM model. Interestingly, upon this model interpretation all the error bars decrease significantly by up to $50\%$, only the error on $m$ remains stable. This also occurs in the SIM-like case.

To quantify the apparent bias, for the two DATA-like cases we compare the corresponding $\chi^2$-values divided by the number of degrees of freedom $\mathrm{ndf} = (N_\mathrm{data}-N_\mathrm{params})$ at the model-independent ShapeFit bestfit (SF) and at the $\Lambda$CDM bestfit (SF$+\Lambda$CDM):
\begin{equation}
\begin{aligned}
    \mathrm{DATAlike~MAX:} \quad &  \chi_\mathrm{SF}^2/\mathrm{ndf} = 62.2/(39-10) \quad & \chi_\mathrm{SF+\Lambda CDM}^2/\mathrm{ndf} = 80.7/(39-6)
    \\
    \mathrm{DATAlike~MIN:} \quad & \chi_\mathrm{SF}^2/\mathrm{ndf} = 63.5/(39-8) \quad & \chi_\mathrm{SF+\Lambda CDM}^2/\mathrm{ndf} = 79.7/(39-4)
\end{aligned}
\end{equation}
Note that the number of free parameters $N_\mathrm{params}$ is reduced by $4$ for the $\Lambda$CDM case, as we fix the four physical parameters to the model prediction when evaluating the corresponding $\chi^2$. As argued before and shown in section \ref{sec:additional-tests}, these relatively poor $\chi^2/\mathrm{ndf}$ of order $\sim 2-2.5$ are mostly induced by the hexadecapole.

For comparison, for the SIMlike case (without hexadecapole) we find a value $\chi_\mathrm{SF}^2/\mathrm{ndf}=11.3/(24-10)$ of order $\sim 1$ similar to other PT challenge participants \cite{Nishimichi:2020tvu}.

This indicates that indeed the ShapeFit best fit without the $\Lambda$CDM condition imposed is better than when imposing it. However the  absolute value of the $\chi^2$ is not a good absolute indicator to be used to test cosmology, because the $\chi^2$ is heavily affected by inadequacies of the modeling of the signal, independently of cosmology- as seen in the hexadecapole behaviour.

\begin{figure}[t]
    \centering
    \begin{minipage}{0.56\textwidth}
    \includegraphics[width=\textwidth]{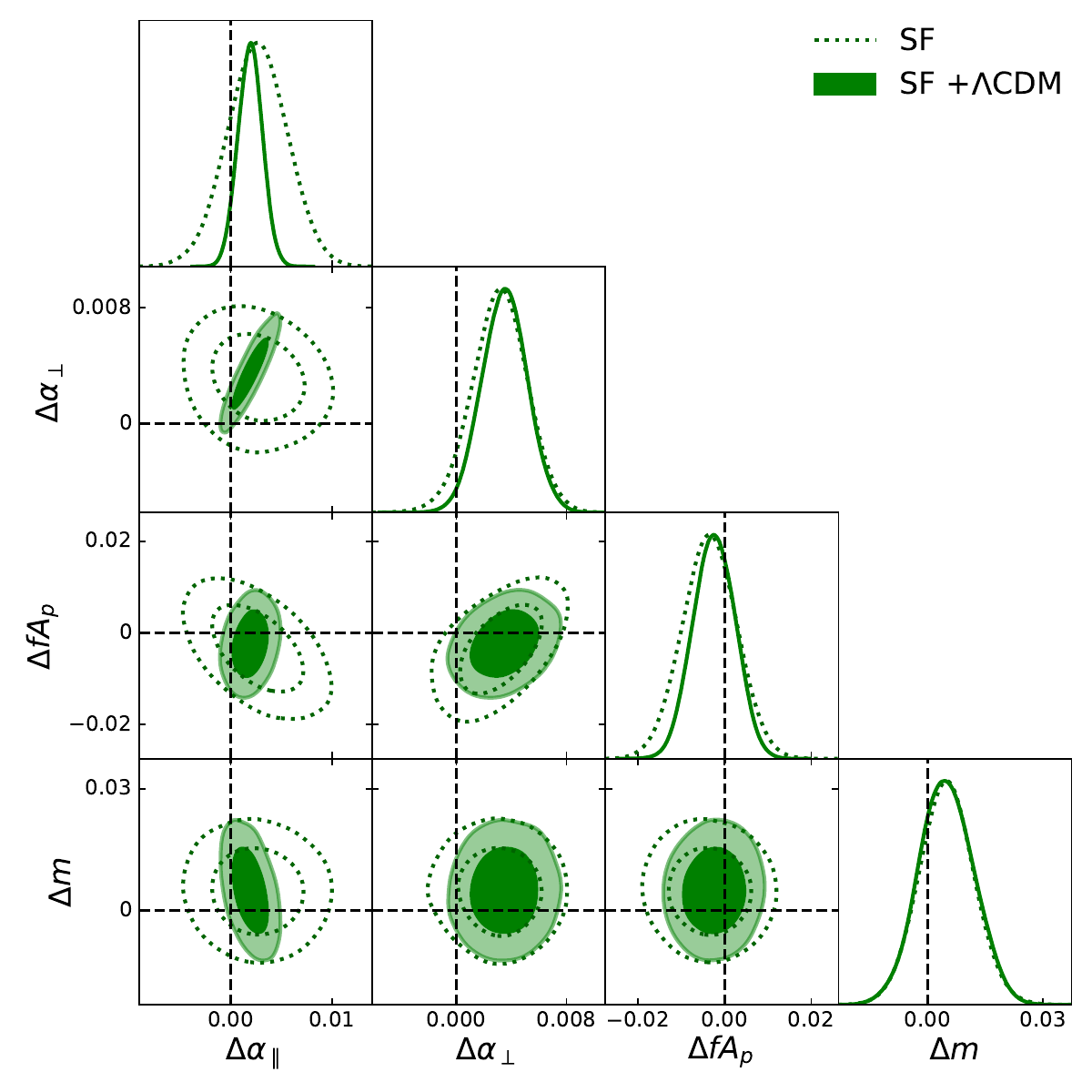}
    \end{minipage}
    \begin{minipage}{0.43\textwidth}
    \vspace{1.8cm}
    \includegraphics[width=\textwidth]{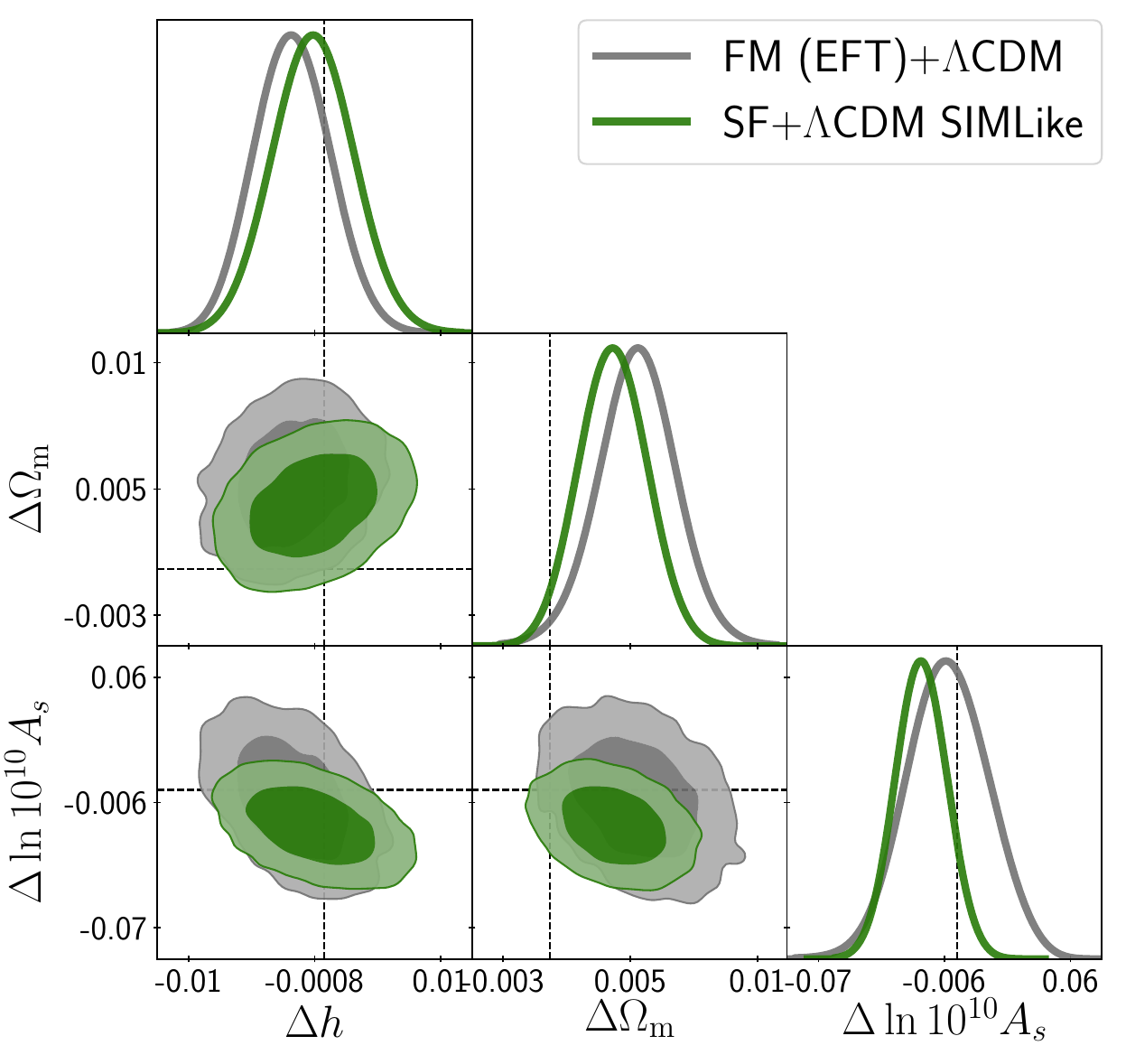}
    \end{minipage}
    \caption{Comparison of our (ShapeFit) main results using the SIM-like convention (green contours, identical to the green contours of Fig.~\ref{fig:main-results}) with the FM case (grey contours) using the EFT implementation of \cite{Ivanov:2019pdj}. In the left panel we show physical parameters constrained via the model-independent ShapeFit (dashed, empty contours) and combined with a $\Lambda$CDM prior (filled contours). The right panel displays the cosmological $\Lambda$CDM parameter space. See text for more discussion.}
    \label{fig:main-results-FM}
\end{figure}

Since the SIM-like case is oriented towards reproducing as much as possible the settings chosen by the  teams participating in \cite{Nishimichi:2020tvu}, here we also show a direct comparison to one of the EFT implementations. In particular, we choose the publicly available CLASS-PT implementation by \cite{Ivanov:2019pdj} and run the EFT model\footnote{When running the EFT model the ShapeFit pipeline and results of the SIM-like configuration were frozen.}  using the same cosmological parameters and priors as in table \ref{tab:methodology-priors}. For the EFT nuisance parameters (bias parameters and counterterms) we choose the default configuration of \cite{Ivanov:2019pdj} but without shot noise correction (i.e., shot noise correction set to zero as shot noise is assumed to be fully under control and correctly subtracted).

The results are shown in figure \ref{fig:main-results-FM}, where the green contours are the same as the SIM-like case in figure \ref{fig:main-results}. Again, we show the physical parameters -with (filled contours) and without (dashed, empty contours) the $\Lambda$CDM prior- in the left panel and the $\Lambda$CDM cosmological parameters in the right panel. As we can see, the SIM-like results are in excellent agreement with the EFT results (grey contours)- both in terms of mean parameters and errors. The ShapeFit SIM-like case errors are somewhat smaller than for EFT,  especially for the amplitude parameter, which might be due to the exclusion of additional significant freedom in extra nuisance parameters, the so-called counterterms, in our model. This is further explored in an upcoming work \cite{Briedendatapaperinprep}. 

\section{Additional tests} \label{sec:additional-tests}

\begin{figure}[t]
\label{fig:main-results-tests1}
    \centering
    \begin{minipage}{0.56\textwidth}
    \includegraphics[width=\textwidth]{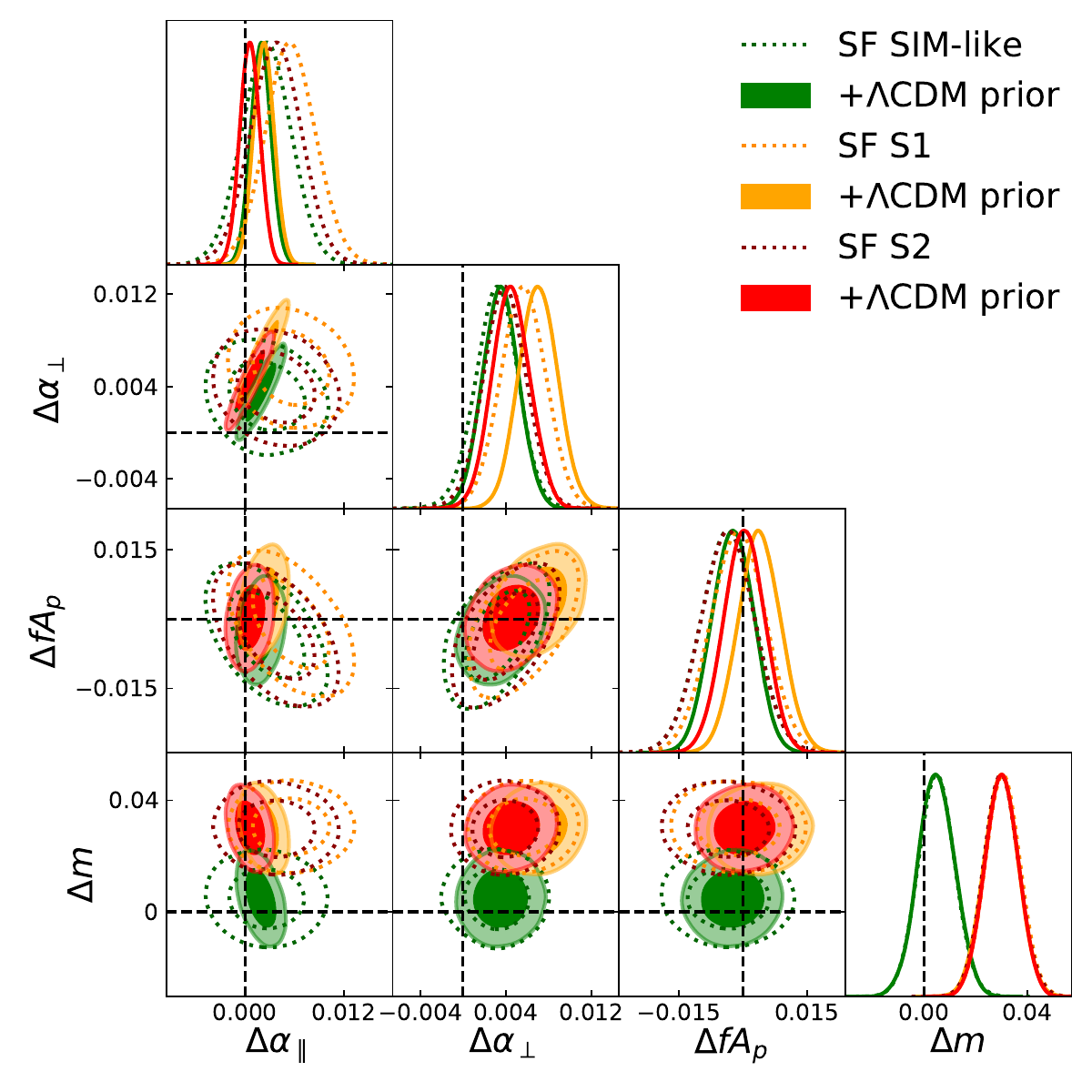}
    \end{minipage}
    \begin{minipage}{0.43\textwidth}
    \vspace{1.8cm}
    \includegraphics[width=\textwidth]{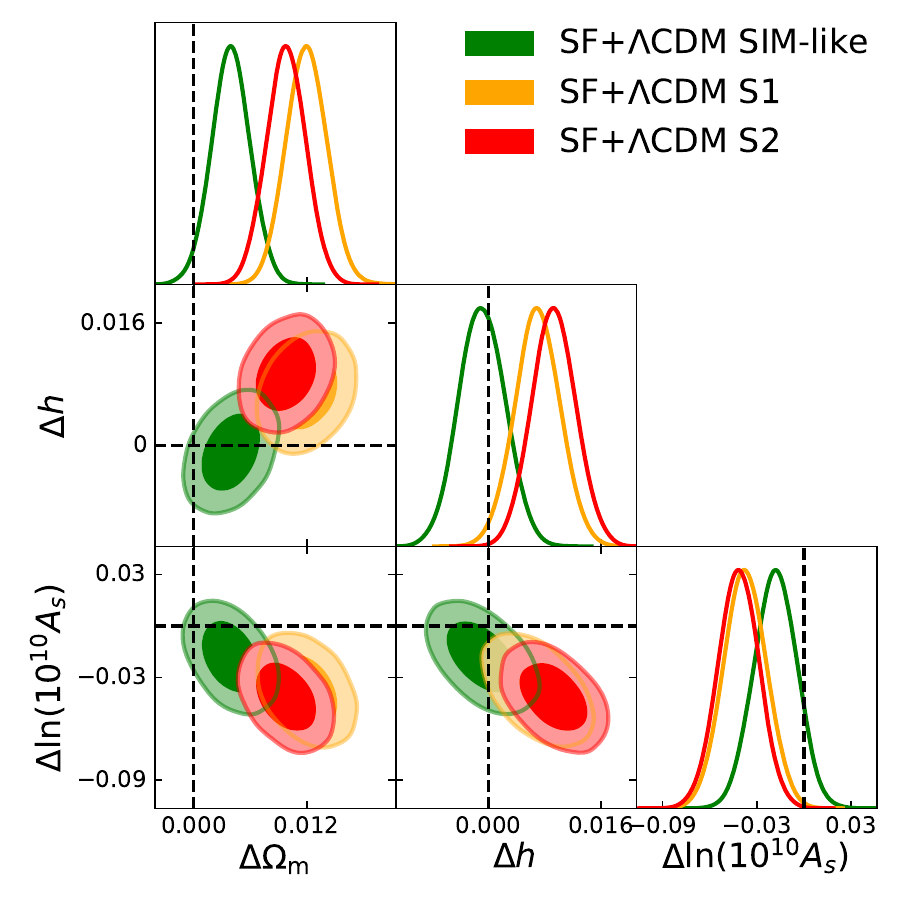}
    \end{minipage}
    \caption{
    Left panel: comparison of the ShapeFit compressed variables either with (solid) or without (dotted) the $\Lambda$CDM prior between the baseline (SIM-like, green) and two analysis variations: in the first we do not apply the geometric correction (S1, orange) and in the second we omit the geometric correction \textit{and} include the $2^\mathrm{nd}$ order PT term. All cases have a maximum wave-vector of $0.12 \hoverMpc$. Right panel: effect on the corresponding cosmological parameters within $\Lambda$CDM model. 
    }
    \label{fig:main-results-tests1}
\end{figure}

As anticipated in table \ref{tab:methodology_all_set-ups} we perform a suite of tests to the blind PT challenge data to quantify the effect of our different model ingredients on the inferred parameters. 
These tests include the impact of i) the geometric correction, ii) including the second order loop correction, iii) dialing the allowed prior range on the shot noise amplitude, iv) choosing a certain bias prescription, v) varying the maximum wave-number used for the analysis, $k_\mathrm{max}$ and vi) including or not the hexadecapole. A few representative cases are illustrated in figures \ref{fig:main-results-tests1} to \ref{fig:main-results-tests-kmax}.

Figure \ref{fig:main-results-tests1} addresses points i) and ii); it shows that the effect of including second-order loop corrections is completely negligible, even for a large survey volume of $566\Gpcoverh^3$. The central values for the parameters are affected at the sub-percent level and the error bars are unaffected. On the other hand, the effect of the geometric correction is very important, but only for $m$, which induces a shift in $m$ of $2\sigma$ when is not taken into account. As expected, the other parameters are relatively unaffected by this correction. This effect is driven by the largest scales: when excluding the large scales by increasing $k_\mathrm{min}$ to $0.02\hoverMpc$ we find that the effect of the geometric correction is reduced by a factor 5 (towards $\Delta m \sim \mathcal{O}(10^{-3})$),
which is of same order of magnitude as the $1\sigma$ precision we find for $m$.
\begin{figure}
    \centering
    \begin{minipage}{0.56\textwidth}
    \includegraphics[width=\textwidth]{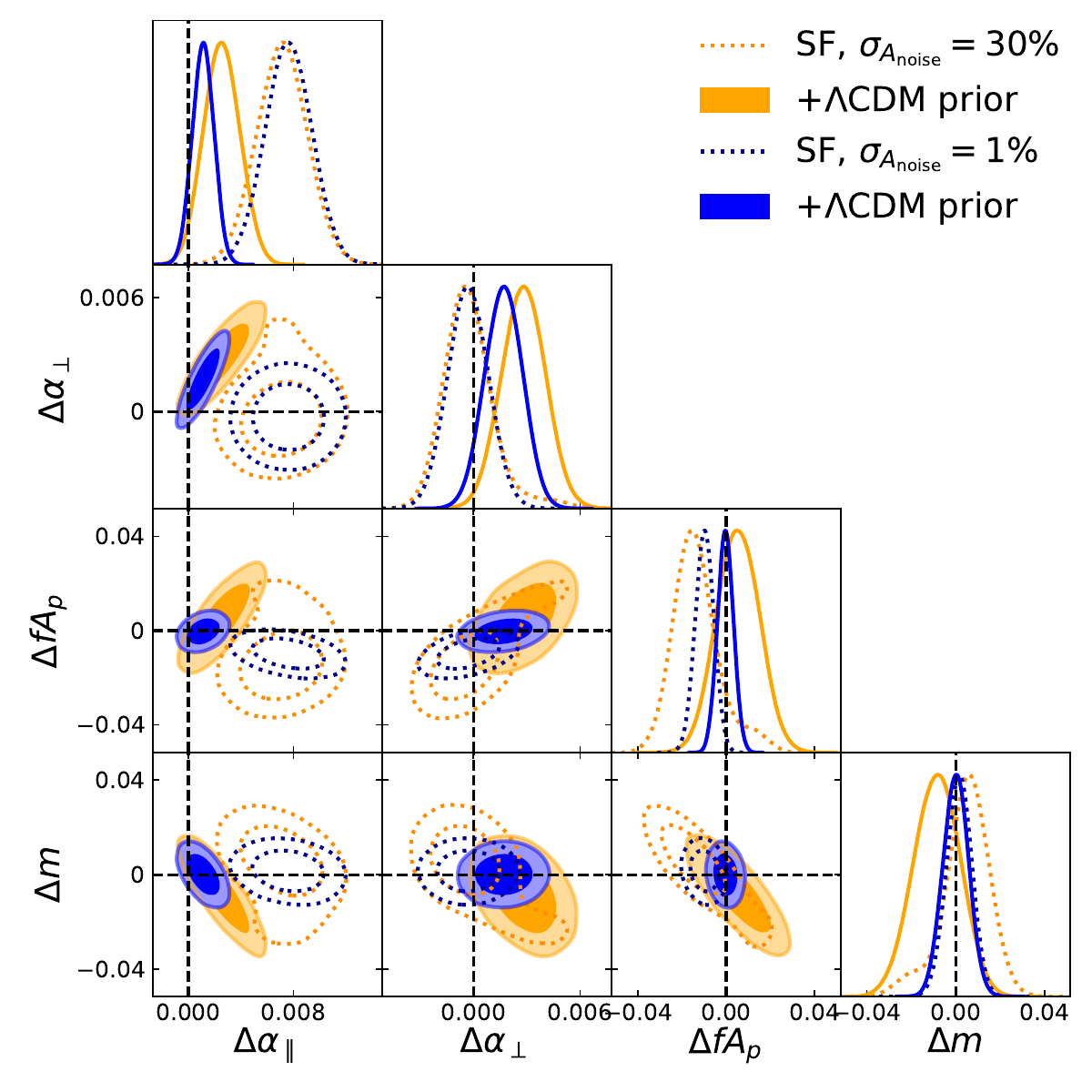}
    \end{minipage}
    \begin{minipage}{0.43\textwidth}
    \vspace{1.8cm}
    \includegraphics[width=\textwidth]{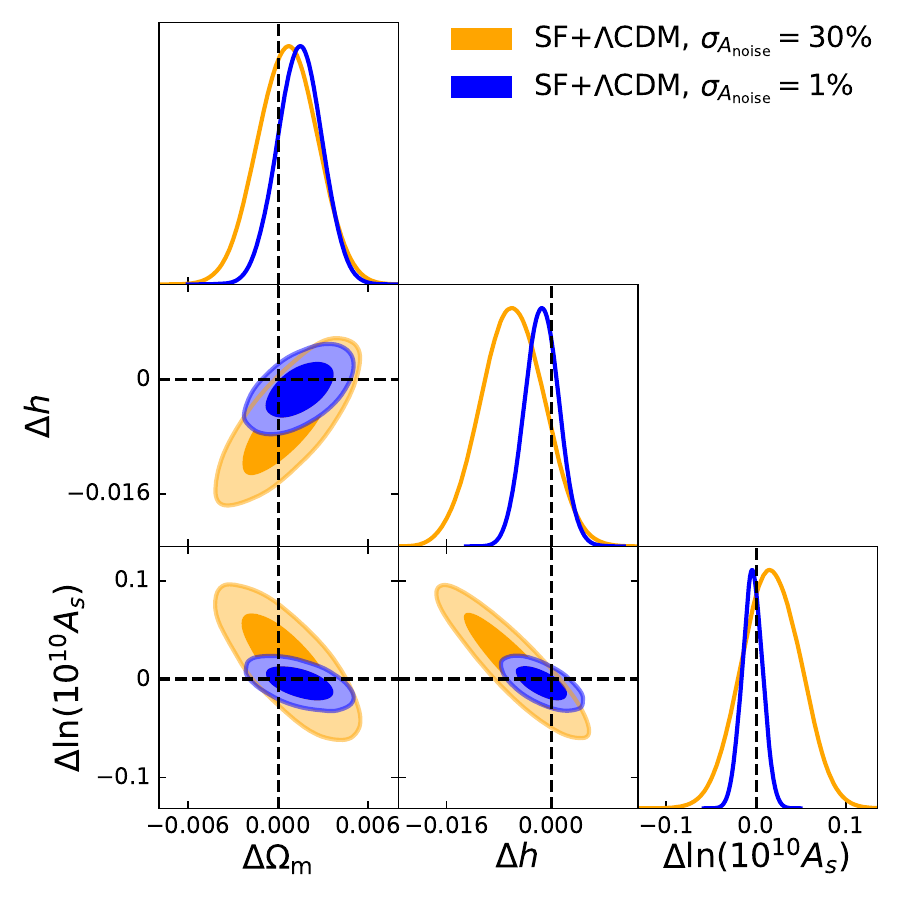}
    \end{minipage}
    \caption{Impact of varying the shot noise term $A_\mathrm{noise}$ within $30\%$ using a Gaussian prior (orange contours) versus effectively fixing it by employing a $1\%$ prior (blue contours). In both cases we allow all bias parameters to vary freely and the fitted $k$-range is $[0.02,0.15]$ using 2LRPT. Compressed variables or shown in the left and cosmological parameters in the right panel.}
    \label{fig:main-results-tests-SN}
\end{figure}

Figure \ref{fig:main-results-tests-SN} displays the impact of varying the amplitude of shot noise as a free parameter. Blue contours show the Poisson prediction with a Gaussian prior of 1\% standard deviation, whereas for the orange contour this is increased up to 30\%. When the shot noise prior is relaxed from 1\% to 30\% strong degeneracies appear with other bias parameters, mostly $b_{s^2}$ and $\sigma_\mathrm{FoG}$, which make the multi-dimensional posteriors highly non-Gaussian. As a consequence, the marginalized low-dimensional posteriors become moderately shifted, and the constraints on cosmological parameters degrade significantly, as shown. In terms of cosmology this degradation in statistical precision highly impacts $h$ and $A_s$, and mildly $\Omega_m$. Thus, we conclude that the choice of the prior around the shot noise amplitude may not severely impact the inferred means of cosmology, but their errors, which in addition tend to become less Gaussian.

\begin{figure}
    \centering
    \begin{minipage}{0.56\textwidth}
    \includegraphics[width=\textwidth]{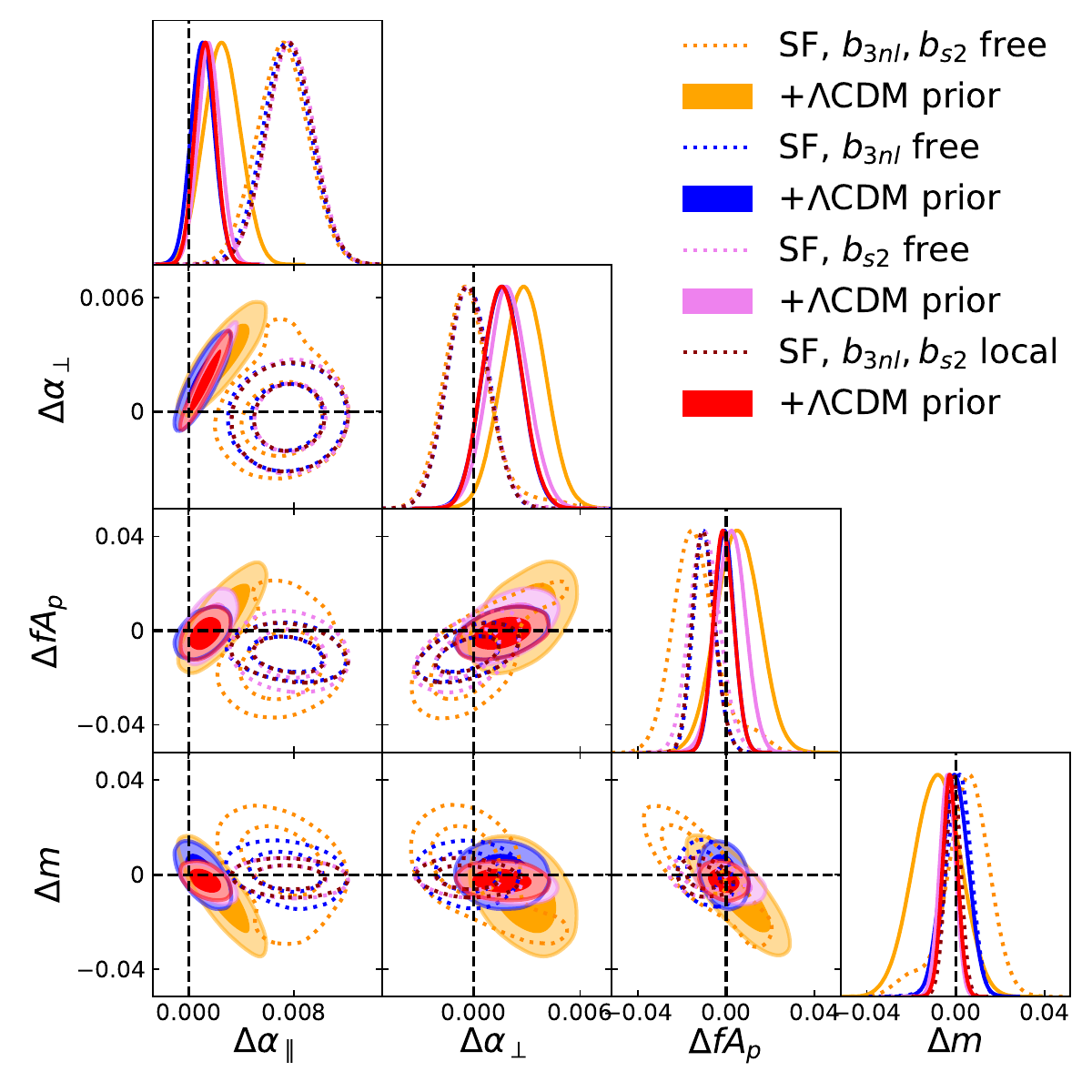}
    \end{minipage}
    \begin{minipage}{0.43\textwidth}
    \vspace{1.8cm}
    \includegraphics[width=\textwidth]{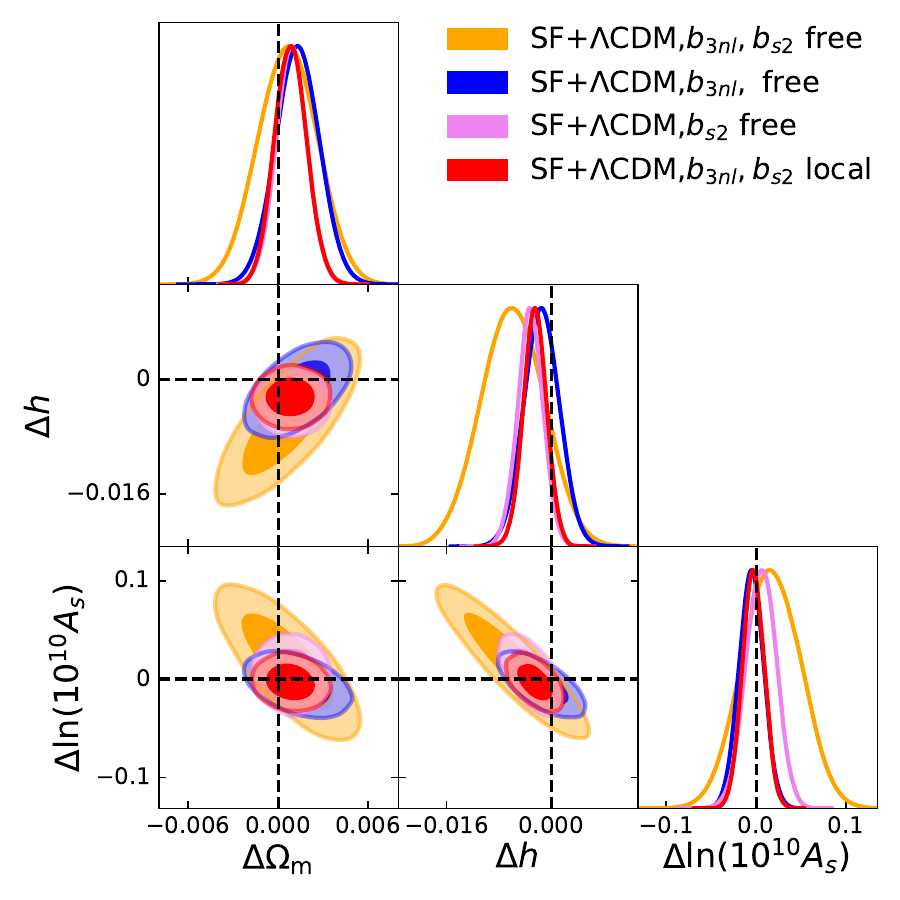}
    \end{minipage}
    \caption{This figure shows the impact of varying the non-local bias parameters $\bnl$ and $\bs$ freely (orange contours) versus setting one of them ($b_{s2}$, blue contours; $b_{3nl}$, violet contours) or both (red contours) to their local Lagrangian prediction. In the left panel we show the compressed parameter constraints for the model-independent ShapeFit (empty, dotted contours) and for the $\Lambda$CDM prior (filled contours).
    }
    \label{fig:main-results-tests-bias}
\end{figure}

The assumptions we make on the non-local bias parameters can highly impact the results on the compressed cosmological parameters. 
In particular, we find that the non-local bias parameter $b_{3\mathrm{nl}}$ is very degenerate with $m$, and therefore using an incorrect value (or functional form with ($b_1-1$) may translate into a systematic shift on our inferred cosmology. When we enforce the local Lagrangian bias relation, $b_{3\mathrm{nl}}=32/315(b_1-1)$ we find that $m$ is biased by $\sim 0.015 \Delta m$ (corresponding to $\sim 1.3\sigma_m$ deviation) with respect to the best-fitting value when $b_{3\mathrm{nl}}$ is allowed to freely vary. We find that this is much less a problem for $b_{s^2}$ than for $b_{3\mathrm{nl}}$. Inevitably, when both biases are free to vary, the error bars on cosmological parameters increase by a factor $2-3$ due to correlations. This is illustrated in figure \ref{fig:main-results-tests-bias} for some representative cases.  
We conclude that for this particular sample, the non-local Lagrangian biases follow the halo local Lagrangian predictions well. Consequently, the main cosmological results are not systematically shifted when relaxing this assumption. However, the resulting error-bars do increase significantly. When the total volume of our sample is closer to that of forthcoming  data (typically 10 to 100 times smaller than the PT challenge sample) this can become a problem, as relaxing both non-local bias parameters significantly weakens or,  in the case of high noise samples, essentially erases the constrains on $m$ (as will be shown in a separate paper \cite{Briedendatapaperinprep}). In these cases, using the local Lagrangian halo relation helps to constrain $m$, and hence use the shape of the power spectrum for cosmology constrains, but at the expenses of being more model-dependent. Alternatively one could impose strong Gaussian priors on higher order bias parameters, as done in most FM analyses, that would otherwise loose a lot of their constraining power. However, from the PT challenge simulations we do not find any strong reasons why to abandon the local Lagrangian assumption.

\begin{figure}
    \centering
    \begin{minipage}{0.50\textwidth}
    \includegraphics[width=\textwidth]{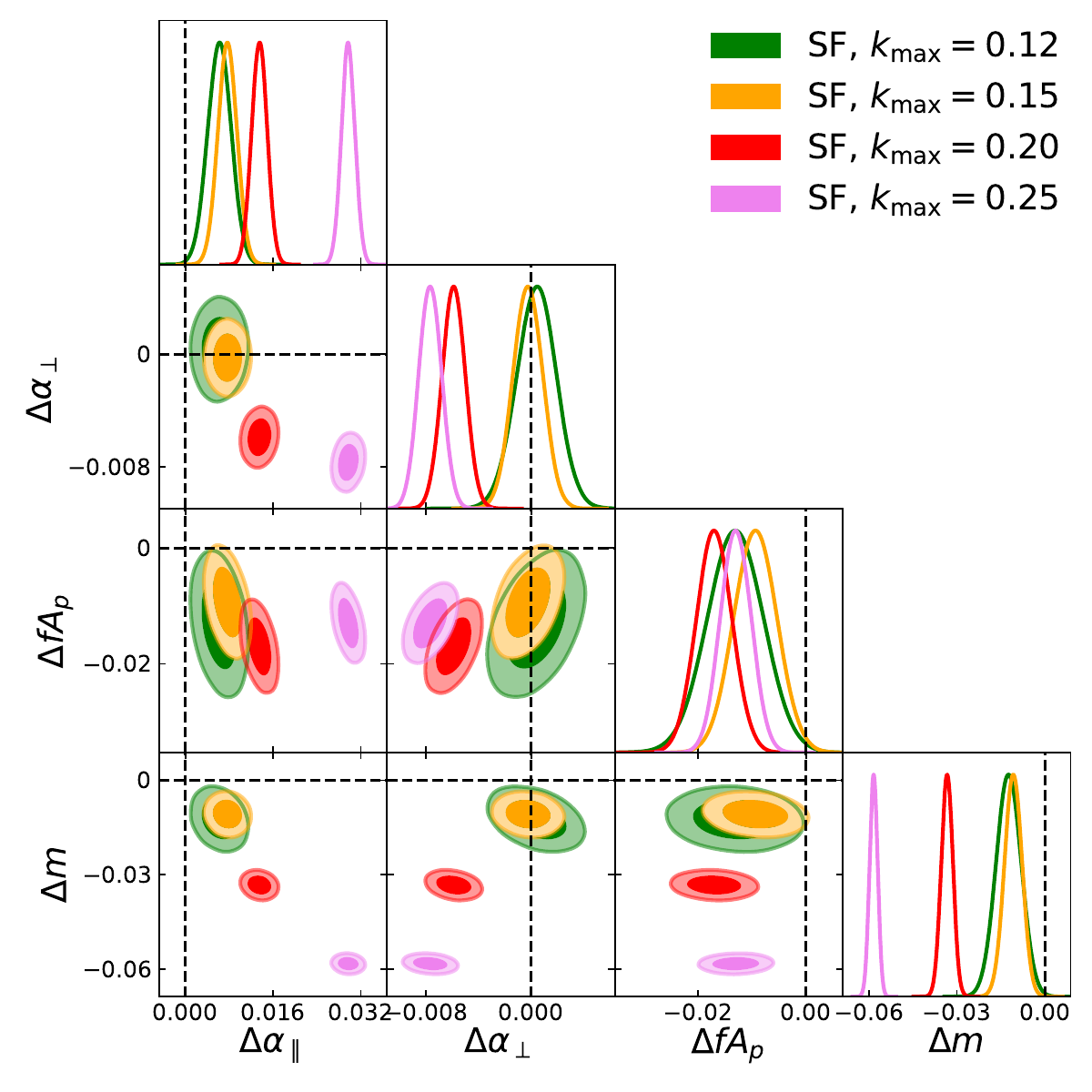}
    \end{minipage}
    \begin{minipage}{0.49\textwidth}
    \includegraphics[width=\textwidth]{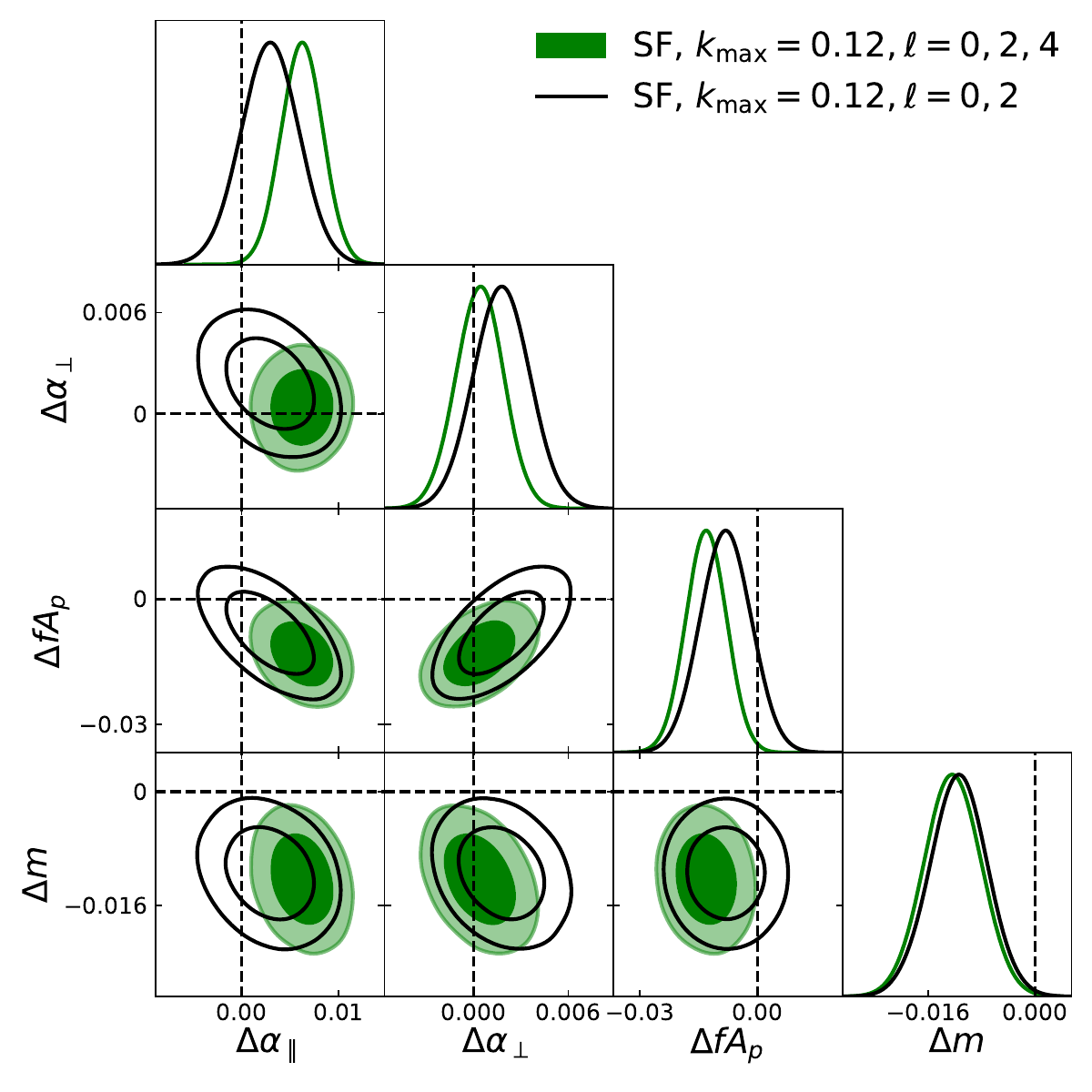}
    \end{minipage}
    \caption{Left panel: Comparison of the ShapeFit compressed variable constraints for different maximum scale cuts, including (in units $\hoverMpc$) $k_\mathrm{max} \in \left\lbrace 0.12, 0.15, 0.20, 0.25 \right\rbrace$ shown in (green, orange, red, violet) filled contours respectively. In all cases we set the minimum scale to $k_\mathrm{min} = 0.02$, the model to 2LRPT, the bias parameters to follow the local Lagrangian prediction and the shot noise amplitude prior to $1\%$. Due to the systematic bias for $k_\mathrm{max} \geq 0.20 \hoverMpc$, we do not show the cosmological $\Lambda$CDM constraints here. Right panel: Comparison of the compressed variables between including (filled, green contours) or not (empty, black contours) the hexadecapole ($\ell = 4$) from the fit. Regarding the remaining set-up, we choose the same as in the left panel, and set $k_\mathrm{max}=0.12 \hoverMpc$. Note that, therefore, the green contours are identical in both panels.}
    \label{fig:main-results-tests-kmax}
\end{figure}

In addition, we compare the results obtained from different $k$ cuts at small scales. We select $k_\mathrm{max} \in \left\lbrace 0.12, 0.15, 0.20, 0.25 \right\rbrace \hoverMpc$, where in all cases we set the minimum wavenumber to $k_\mathrm{min} = 0.02 \hoverMpc$, the model to 2LRPT, the bias parameters to follow the local Lagrangian prediction and the shot noise amplitude prior to $1\%$. 

In the left panel of figure \ref{fig:main-results-tests-kmax} we show the compressed parameter results for these different $k_\mathrm{max}$ cases. While for $k_\mathrm{max}=0.12 \hoverMpc$ and $k_\mathrm{max}=0.15 \hoverMpc$ only $\alpha_\parallel$ appears slightly biased, the fit degrades significantly and leads to large biases once higher wave-numbers are considered. We conclude that the 2LRPT+TNS fixed template approach with local bias assumption is not accurate enough for describing the non-linear scales of $k\geq0.20 \hoverMpc$ for the statistical precision of the PT challenge suite. 

Finally, in the right panel of figure \ref{fig:main-results-tests-kmax} we show the effect of including (or excluding) the hexadecapole signal $(\ell=4)$ from PT challenge data. We see that the hexadecapole biases $\alpha_\parallel$ by $\Delta \alpha_\parallel = 0.005$ and at the same time reduces the error bar by $\sim 50\%$. As anticipated in section \ref{sec:main_results}, this effect together with the effect of increasing $k_\mathrm{max}$ from $0.12 \hoverMpc$ to $0.15 \hoverMpc$ (left panel of figure \ref{fig:main-results-tests-kmax}) drives the bias in $\alpha_\parallel$ observed in the "DATA-like" cases with respect to the "SIM-like" case.

\section{Combining all redshift bins: Constructing a real data analysis scenario}\label{sec:real-allz}

So far, we have only considered cosmological analyses on individual redshift bins. In practice however, surveys observe dark matter tracers throughout a broad redshift range. It is customary to divide these tracers in several, and usually partially overlapping, redshift bins in order to capture cosmological information related to the evolution of cosmic structure across cosmic ages. In this section we analyze the PT challenge in a very similar way. In the FM approach the cosmological parameters are defined at $z=0$ and therefore the redshift evolution is somewhat predetermined within the chosen model. In the classic (and ShapeFit) approach, the compressed variables are determined independently for each redshift bin, hence it is possible to track their time evolution in a model-independent way. The model dependence enters at the very end, when the compressed variables are interpreted in light of a model of choice.

As a preparatory step, we analyze 
the redshift outputs $z_1$ and $z_2$ in the same way as $z_3$ has been analyzed in section~\ref{sec:main_results}. In all cases considered here our baseline set-up is as follows: we fit the monopole, quadrupole, and hexadecapole data using 2LRPT on a wavevector range $0.02 \hoverMpc <k<0.15 \hoverMpc$, where the non-local bias parameters are fixed to their local Lagrangian prediction and the shot noise term is varied within a 30\% Gaussian prior. This corresponds to the set-up labeled "DATA-like MIN" in table \ref{tab:methodology_all_set-ups}. 

Then, we take the following approach: we compute the average of realizations \#1-3 evaluated at $z_1=0.38$, realizations \# 4-7 at $z_2=0.51$, and realizations \# 8-10 at $z_3 = 0.61$. In this way, we reproduce the behaviour of an actual lightcone, which would capture the full redshift range where the 3 redshift bins are independent from each other. This approach closely emulates a real survey catalogue, where different redshift-bins are treated as independent, as long as they are wide enough.\footnote{This assumption is based on the Ergodic hypothesis, and it breaks down for scales comparable to the size of the bin. For this reason it is usual to split the survey in redshift bins which are overlapping and infer their correlation using mocks, as was done for BOSS DR12.}
Also, we retain the same effective number of sampled modes (because we maintain the same volume) as for the single redshift analysis of section \ref{sec:main_results} and the preparatory step.  While in a real world application, partially overlapping redshift bins would not be independent and would thus have a non-zero covariance, the set-up chosen here enables us to compare more directly the findings with the main results, and offers a more transparent interpretation. 

\begin{figure}
    \centering
    \begin{minipage}{0.56\textwidth}
    \includegraphics[width=\textwidth]{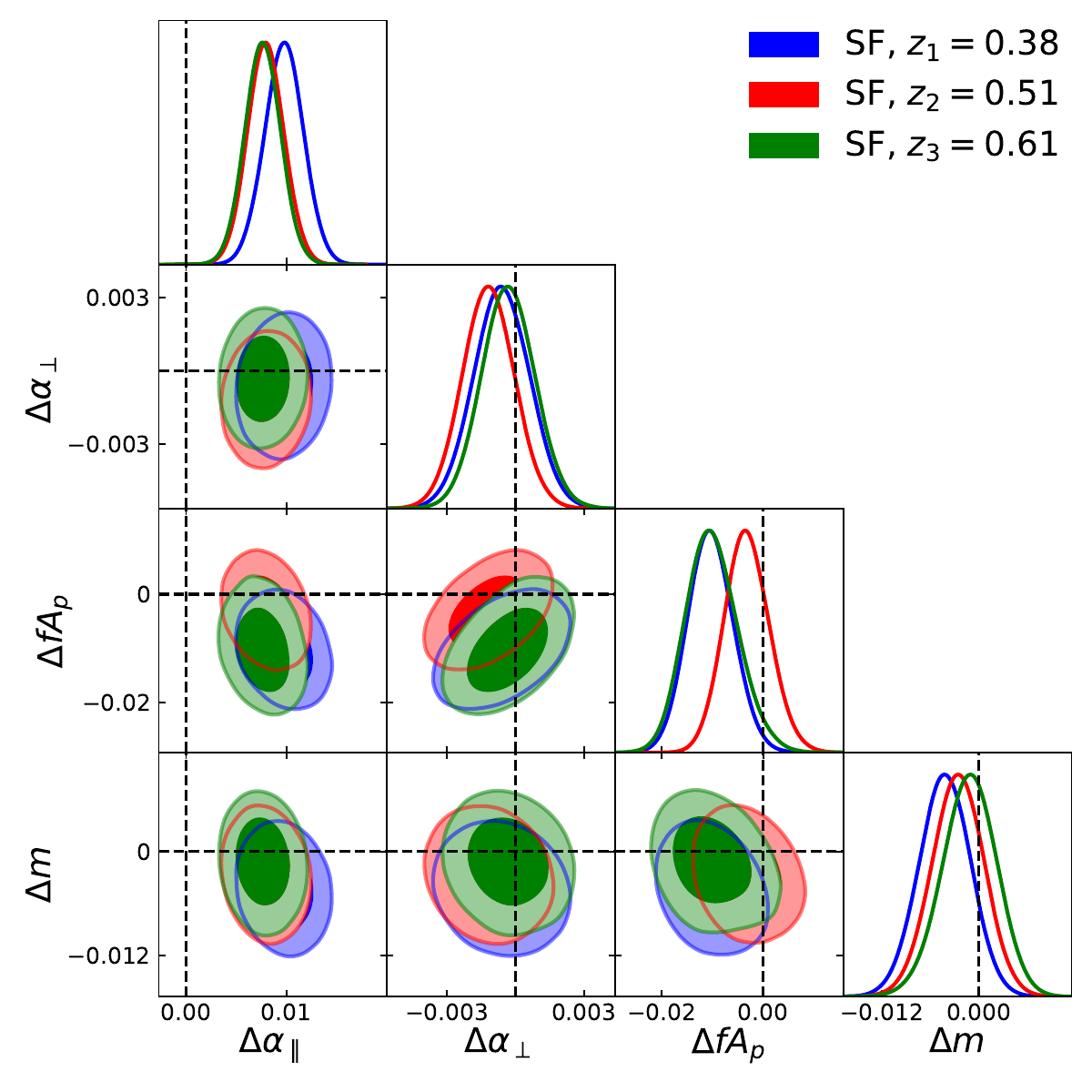}
    \end{minipage}
    \begin{minipage}{0.43\textwidth}
    \vspace{1.8cm}
    \includegraphics[width=\textwidth]{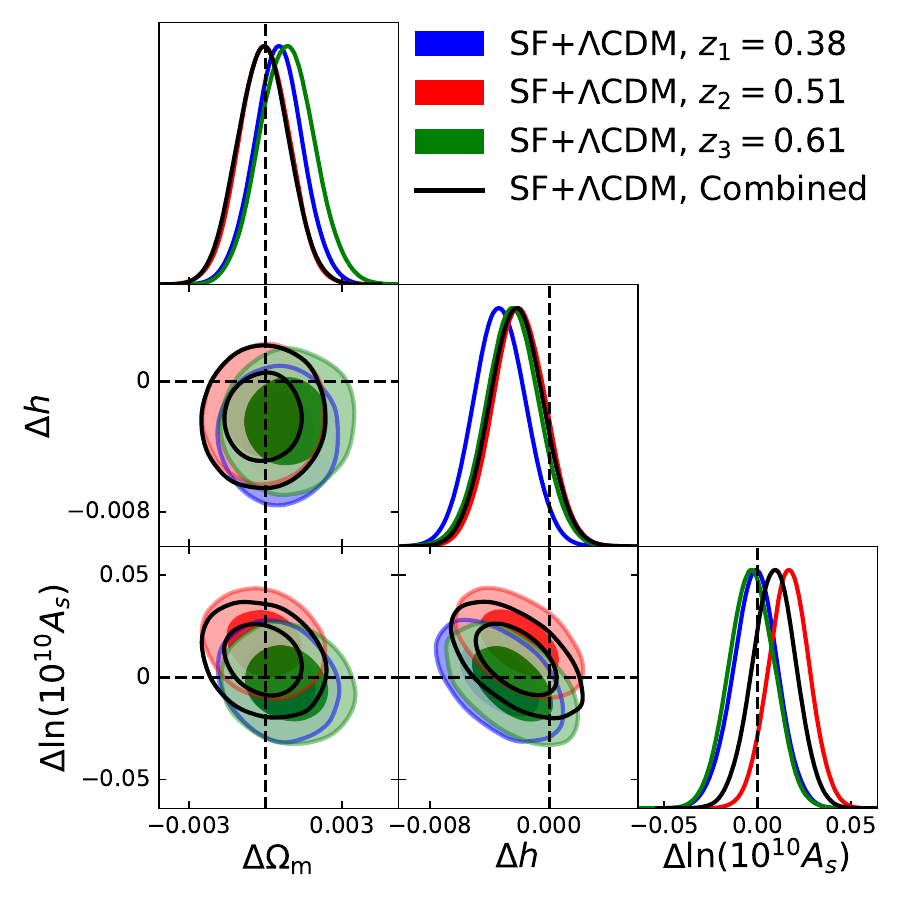}
    \end{minipage}
    \caption{Compressed parameter blind constraints using ShapeFit (left panel) and the corresponding $\Lambda$CDM parameter constraints (right panel) obtained from analysing different redshift bins at $z_1=0.38$ (blue contours), $z_2=0.51$ (red contours) and $z_3=0.61$ (green contours). In addition, we show the cosmological parameter constraints for the case of combining the redshifts, but conserving the total volume, as described in the text (empty black contours). Dashed contours mark the theoretically expected values. The evident bias in $\alpha_\parallel$ is understood  as discussed in section \ref{sec:main_results}  and fig.\ref{fig:main-results}. In all cases we show the "DATA-like MIN" set-up, as customary for a real survey application.}
    \label{fig:main-results-tests-allz}
\end{figure}

In figure \ref{fig:main-results-tests-allz} we show the ShapeFit constraints on compressed variables (left panel) and the cosmological parameter constraints including the $\Lambda$CDM prior (right panel) obtained by analysing the data at different redshifts. On one hand we show the results when analyzing the mean of all 10 realizations evaluated at the same redshift (filled coloured contours). On the other hand, we show the cosmological constraints obtained from the ShapeFit results, for the case when the realizations are spread among the redshifts as described above, including the $\Lambda$CDM prior (empty black contours). Note that all these cases represent constraints based on the same comoving volume, in the coloured cases concentrated in a single redshift bin; and in the empty black contours spread among the 3 redshift bins.  

We see that all redshift bins are consistent with each other within $0.5 \sigma$, both in the compressed parameter and in the cosmological parameter space. The small deviations may arise due to i) differences in non linear structure formation (including non linear galaxy bias and RSD) at different redshifts, both in the N-body simulation and in the fiducial template, ii) different impact of shot noise in the different volumes, iii) and the degree of validity of the assumption of local Lagrangian bias may vary for the different HODs of the different redshift bins.      
As mentioned already in sections \ref{sec:main_results} and \ref{sec:additional-tests}, the ShapeFit constraints on $\alpha_\parallel$ are biased in the "DATA-like MIN" case; this bias vanishes when  applying the $\Lambda$CDM prior. 

One might have expected the combined constraints (empty black contours) on cosmological parameters to be slightly tighter  than the individual redshift constraints, although the effective number of sampled modes is the same. This is  because in the framework of the LCDM model, having information at different epochs (or redshift bins) helps to break degeneracies among model parameters, more effectively than having more volume at a single redshift bin. This effect is seen very clearly, for example, in the $\Omega_m-h$ plane for the BAO cases when Ly-$\alpha$ measurements are combined with low galaxy measurements (see for e.g. fig 1 of \citep{cuceuetal2019}).  Here the effect is not really appreciable because of the smaller redshift range explored. 

\section{Conclusions} \label{sec:conclusion}
We have applied for the first time  the (model-independent)  compressed variables analysis to the (blind) PT challenge simulations. All previous entries to the PT challenge used the full modeling approach instead. This work thus enables a more transparent comparison of these two complementary approaches to the  cosmological analyses and interpretation  of clustering of galaxy redshift surveys.
In particular, for the compressed variables approach we implement ShapeFit \cite{Brieden:2021eduPaper,Brieden:2021cfgLetter} which is fully equivalent to the “classic” approach for the AP and  RSD parameters but includes in addition an extra effective parameter $m$ which captures the power spectrum broadband shape information.

Our chosen baseline (SIM-like) set-up in terms of range of scales, multipoles, shot noise and bias modeling assumptions, prescription for modeling non-linearities, nuisance parameters etc. was chosen to be as close as reasonably possible to the set-up of other PT challenge entries to facilitate comparison.

The volume of the PT challenge simulations is 10 times larger than that envisioned for future surveys, making it possible to uncover small systematic biases even below the level of statistical significance for foreseeable practical applications.  

We find that, in general, ShapeFit recovers the input parameters well within $2\sigma$, and this  accuracy level does not vary dramatically for reasonable changes from the baseline set-up (when changing the range of scales --within reason--,  shot noise and bias modelling assumptions, prescription for modelling non-linearities, nuisance parameters etc.). 

However, we find that the inclusion of the hexadecapole induces a significant systematic shift on the compressed variable  $\alpha_{\parallel}$, when extracted in ShapeFit. This shift would be well below the $2\sigma$ level for future surveys, but clearly indicates that  the modeling of the hexadecapole  should be improved and the effect should be further investigated (for example, it is well-known that the hexadecapole is very sensitive to the inhomogeneous distribution of $k$-vector directions with respect to the line of sight, sec 5.1 of \cite{beutlerboss17}). Nevertheless, it is important to note that this shift is only appreciable in the ShapeFit approach, and the constraints it produces on the model-independent compressed variables. There is no systematic offset in the FM approach within a $\Lambda$CDM model, or when interpreting the constraints on the compressed variables as constraints on the cosmological parameters for a $\Lambda$CDM model.  This is because the systematic shift in $\alpha_{\parallel}$ happens to be in a direction that is not allowed by (or unphysical in) a $\Lambda$CDM model. This teaches us an important lesson:  some systematic biases can be seen at the model-independent compressed variables stage, which would not be spotted in the context of direct, model-dependent fits (especially for minimal-$\Lambda$CDM type-models). More explicitly, it is not sufficient to calibrate and quantify the accuracy of a FM approach on a given family of models (say, $\Lambda$CDM-like models) and then extend it to a different family of models, especially if non-standard, non-trivial extensions of $\Lambda$CDM. Doing so might severely underestimate the  predicted accuracy of the selected FM approach. 

We find that the HOD adopted by the PT challenge produces a galaxy bias which is consistent with the local Lagrangian prescription.  When analyzing real data,  where we have very little control over bias, it may be of interest  to sacrifice precision for accuracy and leave the bias parameters free; this would reduce potential biases at the expense of larger  error-bars.  However, depending on the  survey specifications, leaving the bias parameters completely free might be too conservative,  increasing the error-bars and reducing the signal-to-noise for interesting signatures (see \cite{Briedendatapaperinprep}).

To conclude, looking at the performance of all the PT challenge entries so far (as in \href{https://www2.yukawa.kyoto-u.ac.jp/~takahiro.nishimichi/data/PTchallenge/}{this site}), the agreement is remarkable considering how different the approaches are and that the challenge is blind. Not only the statistical error bars are quite comparable --the statistical error-bars are not blind, but all  approaches  require some compression and compressing can be lossy-- but that the systematic shifts (which are blind) are comparable and under control at least for forthcoming surveys. This agreement is particularly significant when comparing FM with compressed variable approaches given the fundamentally different nature of the two.  
A  direct comparison is presented  in figure \ref{fig:main-results-FM}: our results and constraints are in excellent agreement with the EFT results both in terms of mean parameters and errors. The small residual differences in the size of error-bars which may be due to the priors on specific nuisance parameters, will be explored elsewhere \cite{Briedendatapaperinprep}.

Because of its flexibility, speed, model-independence and, as demonstrated here, precision and accuracy, we envision that the compressed variables approach (including the ShapeFit extension) can offer a valuable contribution in improving the  robustness of  the analysis and interpretation of forthcoming galaxy redshift surveys. 

\acknowledgments
We would like to thank Takahiro Nishimichi for initiating and publishing the blind PT challenge and for useful correspondence regarding our questions about the challenge set-up. Also, we thank Diego Blas for helpful discussions on the IR resummation. H.G-M. and S.B. acknowledges the support from `la Caixa' Foundation (ID100010434) with code LCF/BQ/PI18/11630024. L.V., H.G-M. and S.B. acknowledge support of European Unions Horizon 2020 research and innovation programme ERC (BePreSysE, grant agreement 725327). Funding for this work was partially provided by the Spanish MINECO under projects PGC2018-098866-B-I00 MCIN/AEI/10.13039/501100011033 y FEDER ``Una manera de hacer Europa'', and the ``Center of Excellence Maria de Maeztu 2020-2023'' award to the ICCUB (CEX2019-000918-M funded by MCIN/AEI/10.13039/501100011033)and   MDM-2014-0369 of ICCUB (Unidad de Excelencia Maria de Maeztu). We acknowledge  the IT team at ICCUB
for the help with the Aganice and Hipatia  clusters.

\appendix

\section{Impact of IR resummation correction on ShapeFit}

Since for most FM applications to galaxy power spectra it is crucial to account for large scale bulk flows via the so-called Infrared (IR) resummation, we implement this strategy within our ShapeFit template fits as well and test its impact. Typically, the role of IR resummation is to damp the baryon acoustic oscillation amplitude by an exponential term depending on a certain damping scale $\Sigma(z)$ which depends on redshift. Formally, this damping scale depends on the cosmological model, which is why IR resummation is not usually implemented in the model-independent, ``classic'' template fits. 

Here, we present an approximate scheme, where we ignore this weak cosmology dependence and fix $\Sigma=3.91 \Mpcoverh$, which is the value we obtain for the fiducial template cosmology at $z=0.61$. We adopt the implementation of IR resummation within 1LSPT given by Eq. (7.4) of \citep{blasetal16}, where the corrected, IR resummed non-linear power spectrum is written as a function of the damping scale $\Sigma$, the growth factor $g(z)$ (called $D(z)$ in \citep{blasetal16}) and the linear power spectrum decomposed into a smooth $(P_s)$ and a wiggle $(P_w)$ part. 
The impact of this correction is actually small in the BAO features compared to the standard (No IR) PT, as the exponential factor damping the wiggles gets compensated by the $(1+k^2g^2\Sigma^2)$ factor within the $g^2$ term. Then, the effect of IR in the first loop correction, i.e., the $g^4$ term, is quite small. We do not consider the $g^6$ factor as it represent a small correction in the $k$'s of interest. 

In the left panel of figure \ref{fig:main-results-tests-appendix} we show the impact of including IR resummation (empty, dotted contours) with respect to the baseline (filled contours) on the ShapeFit compressed parameters for either $k_\mathrm{max}=0.12 \hoverMpc$ (green) or $k_\mathrm{max}=0.15 \hoverMpc$ (red). As expected, due to the very small correction of IR resummation on the BAO amplitude, the differences with respect to the baseline are mild. For $k_\mathrm{max} = 0.12$ (green contours) the filled and empty contours are nearly indistinguishable, while for $k_\mathrm{max} = 0.15$ (red contours) we see small differences below $0.2\sigma$ for $\alpha_\parallel, \alpha_\perp, fA_p$ and $1\sigma$ for $m$. However, this is a very mild deviation given the large volume of $566 \Gpcoverh^3$ considered here. In summary the inclusion of IR resummation corrections has no significant effects for the ShapeFit approach.

\section{Impact of the baryon density prior choice} 

Since there might be some confusion in the literature about this, here we test the impact of the prior choice on the baryon density on cosmological constraints. In the case of the PT challenge, the baryon to matter density ratio is known \textit{a priori}, motivated by the baseline $\Lambda$CDM Planck 2018 constraints:

\begin{align}
    \frac{\ob}{\om} = 0.1571~.
\end{align}

However, in most cases in the literature where late time quantities such as the BOSS DR12 and eBOSS DR16 data products are used to constrain cosmological parameters jointly with early time probes, priors on the baryon density today, $\ob$, are adopted. These priors can be motivated either by CMB data, or by the observation of primordial Deuterium and Helium abundances in distant systems, often referred as ``Big Bang Nucleosynthesis'' (BBN) measurements. 

Therefore, it is interesting to investigate to what degree our cosmological parameter constraints depend on the choice of prior information about the baryon density. For that purpose, we choose the bestfit value of $\ob$ from our ``SIM-like'' run as a fiducial value. Then, we perform an additional cosmological fit to the ``SIM-like'' ShapeFit results but using a Gaussian prior centered around that fiducial $\ob$ value with width $\Delta \ob = 0.0001$ instead of imposing a prior on $\ob/\om$ as done in our baseline analysis. The width of the prior is chosen such that $\ob$ is effectively fixed to the same degree as $\ob/\om$ has been fixed for the baseline (see table \ref{tab:methodology-priors}), such that there is no residual effect of the prior when carrying out the comparison.

\begin{figure}
    \centering
    \begin{minipage}{0.56\textwidth}
    \includegraphics[width=\textwidth]{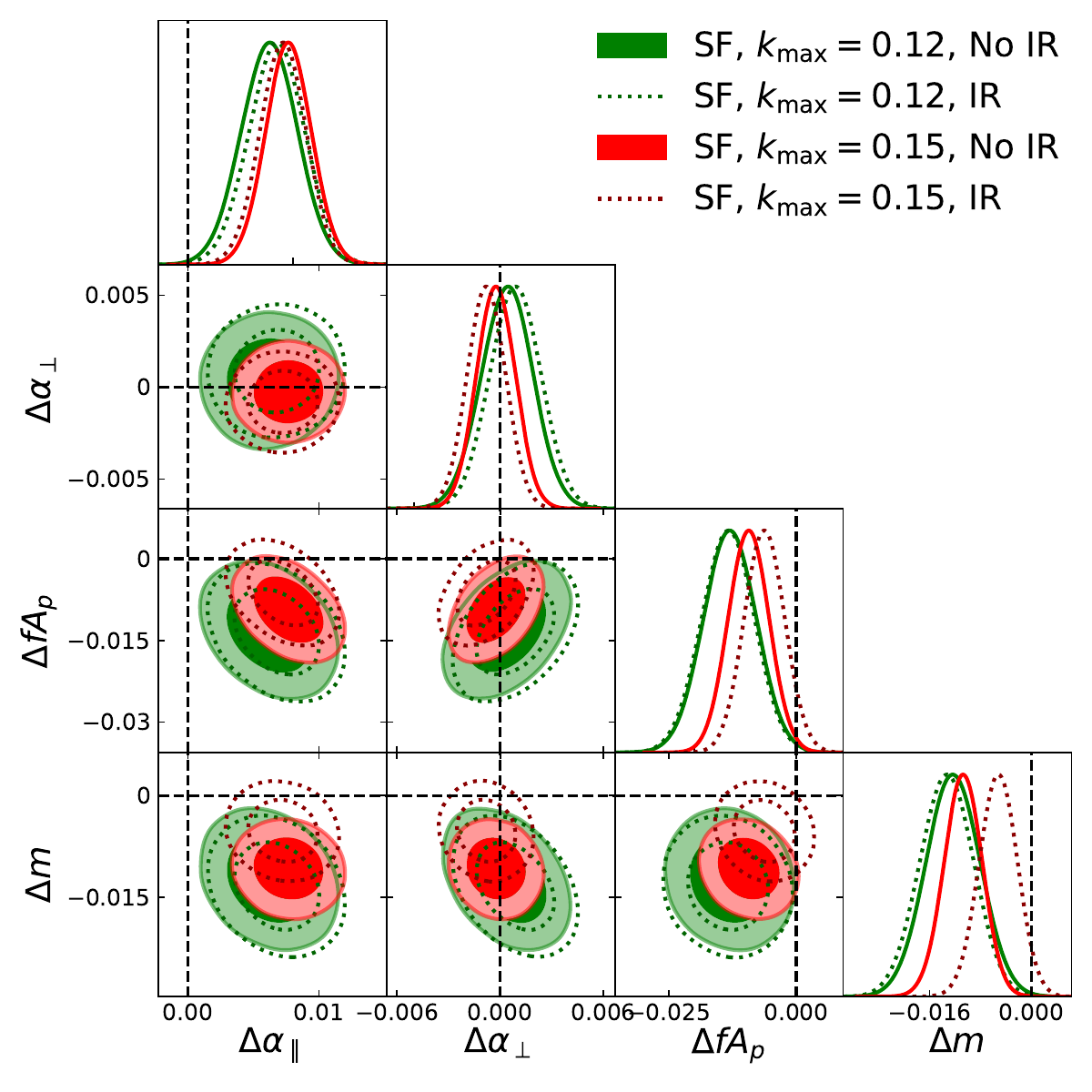}
    \end{minipage}
    \begin{minipage}{0.43\textwidth}
    \vspace{1.8cm}
    \includegraphics[width=\textwidth]{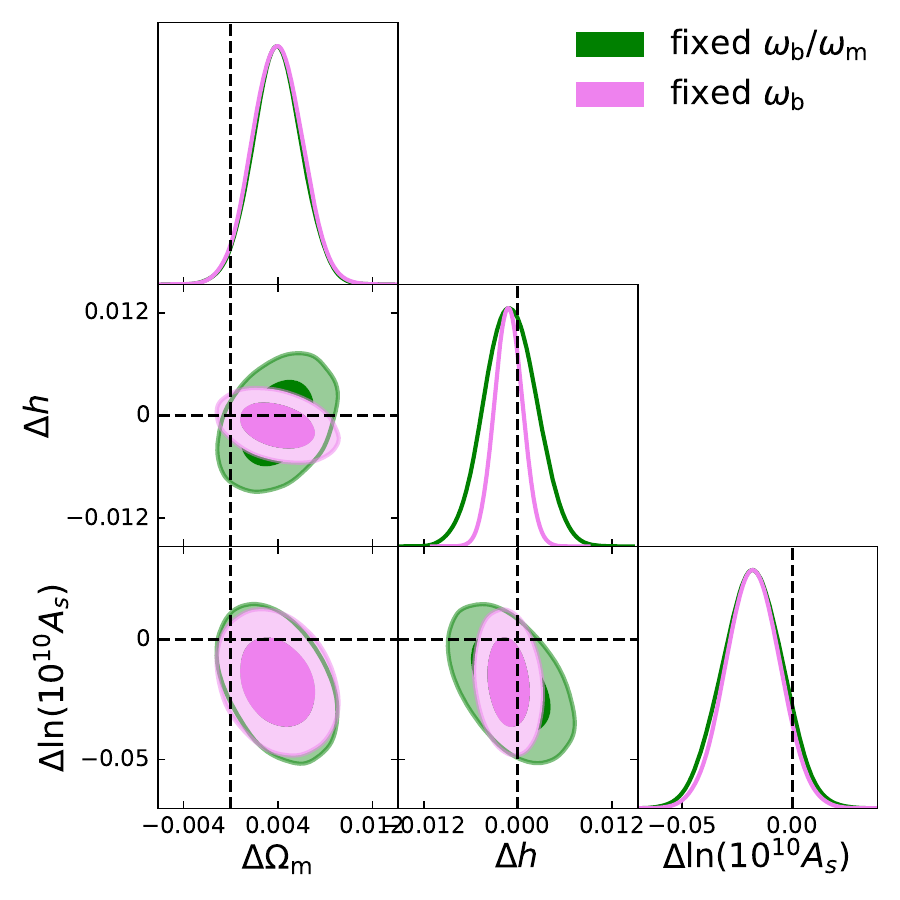}
    \end{minipage}
    \caption{Left panel: Impact on the ShapeFit compressed parameter constraints of including IR resummation (empty, dotted contours) or not (filled contours). The results are presented for two different choices of maximum wavevector, $k_\mathrm{max}=0.12\hoverMpc$ (green) and $k_\mathrm{max}=0.15\hoverMpc$ (red), where the bias parameters follow the local Lagrangian prediction and the shot noise correction prior is fixed to $1\%$. We find that including IR leads to very mild differences typically below $0.2\sigma$ reaching at maximum $1\sigma$ in $m$ for the $k_\mathrm{max}=0.15\hoverMpc$ case. Right panel: blinded cosmological parameters obtained for the "SIM-like" set-up choosing two different implementations of baryon density $\ob$ prior. In green, we show the baseline case for which we use our prior knowledge that the baryon to matter density ratio is set to $\ob/\om = 0.1571$. In violet, we show the results for fixing the baryon density $\ob$ to its best fit value found from the baseline run. The latter choice (motivated by BBN or CMB data) is more often used in the literature when constraining $\Lambda$CDM parameters using actual survey data. We find that the $\Om$ and $A_s$ constraints are insensitive to this prior choice, while constraints on $h$ broaden by a factor $\sim 2$ when choosing a prior on the baryon to matter density ratio rather than the baryon density itself.}
    \label{fig:main-results-tests-appendix}
\end{figure}

The results are presented in the right panel of figure \ref{fig:main-results-tests-appendix}. The baseline ``SIM-like'' results with fixed baryon to matter density ratio are shown in green, while the additional run with fixed baryon density is shown in violet. Of course, both runs deliver the same mean results, as the prior on $\ob$ chosen for the violet contours is given by the bestfit value of the green contours. While the errors on the matter density $\Om$ and the primordial fluctuation amplitude $A_s$ are insensitive to the type of baryon density prior, we see that the constraints on the Hubble parameter $h$ become tighter by a factor 2 once we employ a prior on $\ob$ instead of $\ob/\om$. This is in fact expected, as the former parameter contains information on $h$, while the latter does not.

From this we conclude that the constraints on $\Om$ and $A_s$ from LSS data are more robust and insensitive to the specific prior choice than constraints on $h$. Care has to be taken in particular when showing LSS results on $h$, that made use of an early-time physics assumption, as different, physically well motivated assumptions can lead to subtle differences in the posterior constraints.

%
%
%


\def\jnl@style{\it}
\def\aaref@jnl#1{{\jnl@style#1}}

\def\aaref@jnl#1{{\jnl@style#1}}

\def\aj{\aaref@jnl{AJ}}                   
\def\araa{\aaref@jnl{ARA\&A}}             
\def\apj{\aaref@jnl{ApJ}}                 
\def\apjl{\aaref@jnl{ApJ}}                
\def\apjs{\aaref@jnl{ApJS}}               
\def\ao{\aaref@jnl{Appl.~Opt.}}           
\def\apss{\aaref@jnl{Ap\&SS}}             
\def\aap{\aaref@jnl{A\&A}}                
\def\aapr{\aaref@jnl{A\&A~Rev.}}          
\def\aaps{\aaref@jnl{A\&AS}}              
\def\azh{\aaref@jnl{AZh}}                 
\def\baas{\aaref@jnl{BAAS}}               
\def\jrasc{\aaref@jnl{JRASC}}             
\def\memras{\aaref@jnl{MmRAS}}            
\def\mnras{\aaref@jnl{MNRAS}}             
\def\pra{\aaref@jnl{Phys.~Rev.~A}}        
\def\prb{\aaref@jnl{Phys.~Rev.~B}}        
\def\prc{\aaref@jnl{Phys.~Rev.~C}}        
\def\prd{\aaref@jnl{Phys.~Rev.~D}}        
\def\pre{\aaref@jnl{Phys.~Rev.~E}}        
\def\prl{\aaref@jnl{Phys.~Rev.~Lett.}}    
\def\pasp{\aaref@jnl{PASP}}               
\def\pasj{\aaref@jnl{PASJ}}               
\def\qjras{\aaref@jnl{QJRAS}}             
\def\skytel{\aaref@jnl{S\&T}}             
\def\solphys{\aaref@jnl{Sol.~Phys.}}      
\def\sovast{\aaref@jnl{Soviet~Ast.}}      
\def\ssr{\aaref@jnl{Space~Sci.~Rev.}}     
\def\zap{\aaref@jnl{ZAp}}                 
\def\nat{\aaref@jnl{Nature}}              
\def\iaucirc{\aaref@jnl{IAU~Circ.}}       
\def\aplett{\aaref@jnl{Astrophys.~Lett.}} 
\def\apspr{\aaref@jnl{Astrophys.~Space~Phys.~Res.}}
\def\bain{\aaref@jnl{Bull.~Astron.~Inst.~Netherlands}} 
\def\fcp{\aaref@jnl{Fund.~Cosmic~Phys.}}  
\def\gca{\aaref@jnl{Geochim.~Cosmochim.~Acta}}   
\def\grl{\aaref@jnl{Geophys.~Res.~Lett.}} 
\def\jcp{\aaref@jnl{J.~Chem.~Phys.}}      
\def\jgr{\aaref@jnl{J.~Geophys.~Res.}}    
\def\jqsrt{\aaref@jnl{J.~Quant.~Spec.~Radiat.~Transf.}}
\def\memsai{\aaref@jnl{Mem.~Soc.~Astron.~Italiana}}
\def\nphysa{\aaref@jnl{Nucl.~Phys.~A}}   
\def\physrep{\aaref@jnl{Phys.~Rep.}}   
\def\physscr{\aaref@jnl{Phys.~Scr}}   
\def\planss{\aaref@jnl{Planet.~Space~Sci.}}   
\def\procspie{\aaref@jnl{Proc.~SPIE}}   
\def\jcap{\aaref@jnl{J. Cosmology Astropart. Phys.}}

\let\astap=\aap
\let\apjlett=\apjl
\let\apjsupp=\apjs
\let\applopt=\ao

\newcommand{\etal}{et al.\ }

\newcommand{\mpc}{\, {\rm Mpc}}
\newcommand{\kpc}{\, {\rm kpc}}
\newcommand{\hmpc}{\, h^{-1} \mpc}
\newcommand{\ihmpc}{\, h\, {\rm Mpc}^{-1}}
\newcommand{\ikms}{\, {\rm s\, km}^{-1}}
\newcommand{\kms}{\, {\rm km\, s}^{-1}}
\newcommand{\hkpc}{\, h^{-1} \kpc}
\newcommand{\lya}{Ly$\alpha$\ }
\newcommand{\lyb}{Lyman-$\beta$\ }
\newcommand{\lyaf}{Ly$\alpha$ forest}
\newcommand{\lr}{\lambda_{{\rm rest}}}
\newcommand{\bF}{\bar{F}}
\newcommand{\bS}{\bar{S}}
\newcommand{\bC}{\bar{C}}
\newcommand{\bB}{\bar{B}}
\newcommand{\vdF}{{\mathbf \delta_F}}
\newcommand{\vdS}{{\mathbf \delta_S}}
\newcommand{\vdf}{{\mathbf \delta_f}}
\newcommand{\vdn}{{\mathbf \delta_n}}
\newcommand{\vdC}{{\mathbf \delta_C}}
\newcommand{\vdX}{{\mathbf \delta_X}}
\newcommand{\xrei}{x_{rei}}
\newcommand{\lrmin}{\lambda_{{\rm rest, min}}}
\newcommand{\lrmax}{\lambda_{{\rm rest, max}}}
\newcommand{\lmin}{\lambda_{{\rm min}}}
\newcommand{\lmax}{\lambda_{{\rm max}}}
\newcommand{\hi}{\mbox{H\,{\scriptsize I}\ }}
\newcommand{\heii}{\mbox{He\,{\scriptsize II}\ }}
\newcommand{\vp}{\mathbf{p}}
\newcommand{\vq}{\mathbf{q}}
\newcommand{\vxperp}{\mathbf{x_\perp}}
\newcommand{\vkperp}{\mathbf{k_\perp}}
\newcommand{\vrperp}{\mathbf{r_\perp}}
\newcommand{\vx}{\mathbf{x}}
\newcommand{\vy}{\mathbf{y}}
\newcommand{\vk}{\mathbf{k}}
\newcommand{\vR}{\mathbf{r}}
\newcommand{\tdtwo}{\tilde{b}_{\delta^2}}
\newcommand{\tstwo}{\tilde{b}_{s^2}}
\newcommand{\tbthree}{\tilde{b}_3}
\newcommand{\tadtwo}{\tilde{a}_{\delta^2}}
\newcommand{\tastwo}{\tilde{a}_{s^2}}
\newcommand{\tabthree}{\tilde{a}_3}
\newcommand{\tpsi}{\tilde{\psi}}
\newcommand{\vv}{\mathbf{v}}
\newcommand{\fnl}{{f_{\rm NL}}}
\newcommand{\tfnl}{{\tilde{f}_{\rm NL}}}
\newcommand{\gnl}{g_{\rm NL}}
\newcommand{\orderfour}{\mathcal{O}\left(\delta_1^4\right)}
\newcommand{\SDSSPF}{\cite{2006ApJS..163...80M}}
\newcommand{\PF}{$P_F^{\rm 1D}(k_\parallel,z)$}
\newcommand\ionalt[2]{#1$\;${\scriptsize \uppercase\expandafter{\romannumeral #2}}}%
\newcommand{\vxone}{\mathbf{x_1}}
\newcommand{\vxtwo}{\mathbf{x_2}}
\newcommand{\vRot}{\mathbf{r_{12}}}
\newcommand{\cm}{\, {\rm cm}}


\providecommand{\href}[2]{#2}\begingroup\raggedright\endgroup

\end{document}